\newcommand{\arxiv}[1]{}
\newcommand{\oops}[1]{}
\newcommand{\iclp}[1]{#1}      
\newcommand{\noticlp}[1]{}     
\newcommand{\iclparxiv}[2]{#1} 

\noticlp{
\documentclass[12pt]{article}
\usepackage{fullpage}
}

\iclp{
\documentclass{new_tlp}
}

\oops{
\PassOptionsToPackage{dvipsnames}{xcolor}
\documentclass[acmsmall,anonymous,review,screen]{acmart}

\acmConference[POPL '23]{}{June 03--05,
  2018}{Woodstock, NY}
} 

\usepackage{comment}
\newif\ifoopsenv
\oopsenvfalse  
\ifoopsenv
  
\else
  \excludecomment{oopsenv}
\fi

\newif\ifarxivenv
\arxivenvtrue    
\ifarxivenv
  
\else
  \excludecomment{arxivenv}
\fi

\usepackage{float}

\usepackage{hyperref} 
\usepackage{xspace}
\usepackage{alltt}
\usepackage[T1]{fontenc}

\usepackage{enumitem}
\setlist[description]{itemindent=-2ex}
\oops{
\setlist{leftmargin=3ex, itemsep=.6ex} 
}
\iclp{
\setlist{leftmargin=3ex, itemsep=0ex, topsep=0.5ex} 
}

\usepackage[dvipsnames]{xcolor}
\definecolor{codegray}{rgb}{0.5,0.5,0.5}
\definecolor{codepurple}{rgb}{0.58,0,0.82}

\usepackage{amsmath,amssymb}
\usepackage{graphicx}
\usepackage{multirow}
\usepackage{multicol}
\usepackage{setspace} 

\noticlp{
\usepackage{tikz}
\usepackage{pgfplots}
\usepackage{csvsimple}
\usepackage{filecontents}
\usepackage{pgfplotstable} 
\pgfplotsset{compat=1.9} 
\pgfplotsset{compat=1.9, compat/bar nodes=1.8, }
\pgfplotsset{compat=1.15}
}

\usepackage{listings}

\lstdefinestyle{mystyle}{
    backgroundcolor=\color{white},   
    basicstyle=\noticlp{\scriptsize}\iclp{\footnotesize}\ttfamily,
    commentstyle=\color{brown},
    keywordstyle=\color{blue}, 
    stringstyle=\color{codepurple},
    numberstyle=\tiny\color{codegray},
    numbers=none,
    breakatwhitespace=false,
    breaklines=false,                 
    captionpos=b,
    keepspaces=true,                 
    numbersep=5pt,
    showspaces=false,                
    showstringspaces=false,
    showtabs=false,                  
    tabsize=2
}
\newcommand{\Hex}[1]{\hspace{#1ex}}
\newcommand{\Vex}[1]{\vspace{#1ex}}
\newcommand{\itempar}{\oops{\Vex{.5}}}

\newenvironment{code}{\Vex{\noticlp{.1}\iclp{-.5}}\begin{alltt}\noticlp{\small}\iclp{\footnotesize}}{\end{alltt}\Vex{\noticlp{.5}\iclp{-1}}}
\newcommand\co[1]{\mbox{\noticlp{\small}\iclp{\footnotesize}\tt #1}} 
\newcommand\kw[1]{\textcolor{blue}{#1}} 
\newcommand{\mathify}[1]{\ifmmode{\mbox{$#1$}}\else\mbox{$#1$}\fi}
\newcommand\m[1]{$#1$} 
\newcommand\p[1]{\mathify{\it #1}} 
\newcommand\bigO[1]{\m{O(#1)}} 

\newcommand\WHILE{\kw{while}\xspace}
\newcommand\INFER{\kw{infer}}
\newcommand\RULES{\kw{rules}}
\newcommand\cIF{\kw{if}\xspace}

\newcommand\NOT{\kw{not}\xspace}
\newcommand\SOME{\kw{some}\xspace}

\newcommand\IN{\kw{in}\xspace}

\newcommand{\extern}{\co{external\_file\_query}\xspace}

\newcommand{\mysec}[1]{\iclp{\Vex{-1}}\section{#1}\iclp{\Vex{-.5}}}
\newcommand{\mysubsec}[1]{\iclp{\Vex{-1.5}}\subsection{#1}\iclp{\Vex{-.5}}}
\newcommand{\mypar}[1]{\Vex{\noticlp{2}\iclp{1.5}} \noindent {\bf #1.~~}}
\newenvironment{myquote}[1]%
  {\list{}{\leftmargin=#1\rightmargin=#1}\item[]}%
  {\endlist}
\newenvironment{example}{\Vex{.5}\par\textbf{\textit{Example}}.}{\hfill$\blacksquare$\par}

\newcommand\defn[1]{{\it #1}}

\newcommand\myurl[1]{{\footnotesize\url{#1}}}
\newcommand\ghurl[2]{} 

\newcommand{\notes}[1]{}

\usepackage{longtable}

\begin{document}

\newcommand\fund{
    This work was supported in part by NSF under grants 
    CCF-1954837, 
    CCF-1414078, and 
    IIS-1447549 
    and ONR under grants
    N00014-21-1-2719, 
    N00014-20-1-2751, and 
    N00014-15-1-2208. 
}

\title{\noticlp{Programming with Logic Rules
    and}\iclp{\m{\!}Integrating~Logic\,Rules~with}~Everything~Else,~Seamlessly\m{\!}\noticlp{\thanks{\fund}}}

\noticlp{
\author
{Yanhong A.\ Liu \Hex{1.5} Scott D.\ Stoller \Hex{1.5} 
  Yi Tong \Hex{1.5} Bo Lin\\
  Stony Brook University\\
  \{liu,stoller,yittong,bolin\}@cs.stonybrook.edu}
\date{July 22, 2022} 
}

\oops{
\author{Yanhong A. Liu}
\email{liu@cs.stonybrook.edu}          
\orcid{nnnn-nnnn-nnnn-nnnn}             
\affiliation{
  \institution{Stony Brook University}            
  \city{Stony Brook}
  \state{New York}
  \country{USA}                    
}
\author{Scott D.\ Stoller}
\author{Yi Tong}
\author{Bo Lin}
\author{K.\ Tuncay Tekle}
}

\iclp{
\author[Y.A.\ Liu, S.D.\ Stoller, Y.\ Tong, and B.\ Lin]
{YANHONG A.\ LIU \Hex{2.5} SCOTT D.\ STOLLER \Hex{2.5} 
  YI TONG \Hex{2.5} BO LIN\\
  Stony Brook University, Stony Brook, NY 11794, USA\\
  \email{\{liu,stoller,yittong,bolin\}@cs.stonybrook.edu}}

\jdate{May 2023}
\pubyear{2023}
\pagerange{\pageref{firstpage}--\pageref{lastpage}}
\submitted{February 2023}
\revised{April 2023}
\accepted{May 2023}

\label{firstpage}
}

\maketitle

\begin{abstract}

\noticlp{
Logic rules are powerful for expressing complex reasoning and analysis
problems.  At the same time, they are inconvenient or impossible to use
for many other aspects of applications.  Integrating logic rules in a language
with sets and functions, and furthermore with updates and objects, has been
a subject of significant study.  What's lacking is a language that
integrates all constructs seamlessly.
}

This paper presents a language, Alda, that supports all of logic rules,
sets, functions, updates, and objects as seamlessly integrated built-ins.
%
The key idea is to support predicates in rules as set-valued variables that
can be used and updated in any scope, and support queries using rules as
either explicit or implicit automatic calls to an inference function.
We have defined a formal semantics of the language,
\notes{
We develop an integrated declarative and operational semantics
that naturally extends and seamlessly combines 
the semantics of all of 
rules, 
sets, 
functions, 
updates, and
objects including concurrent and distributed processes.}%
implemented a prototype compiler that builds on an object-oriented language
that supports concurrent and distributed programming and on an efficient logic rule system,
%
and successfully used the language and implementation on benchmarks and problems
from a wide variety of application domains. 
We describe the compilation method and results of experimental evaluation.

\end{abstract}

\oops{
\maketitle
}

\iclp{
}

\mysec{Introduction}

%
%
%
%

Logic rules are powerful for expressing complex reasoning and analysis
problems, especially in critical areas such as program analysis,
decision support, networking, and
security~\cite{WarLiu17AppLP-arxiv,liu18LPappl-wbook}.  However, developing
application programs that use logic rules remains challenging:
\begin{itemize}

\item Powerful logic languages and systems support succinct use of logic
  rules for complex reasoning and analysis, but not as directly or
  conveniently for many other aspects of applications---e.g., data
  aggregation, numerical computation, input/output, modular construction,
  and concurrency---that are more easily expressed using set queries,
  functions, state updates, and object
  encapsulation~\cite{maier18hist-wbook}.

\item At the same time, commonly-used languages for building applications
  support many powerful features but not logic rules, and to use a logic rule
  system, tedious and error-prone interface code is required---to pass
  rules and data to the rule system, invoke operations of the 
  rule system for answering queries, and pass the results back---
  manually solving an impedance mismatch, similarly as in interfaces with
  relational databases~\cite{geiger1995inside}, making logic rules harder
  to use than necessary.

\end{itemize}
What's lacking is
(1) %
a simple and powerful language that can express application problems
by directly using logic rules as well as all other features
without extra interface code, and with a clear semantics for analysis
as well as execution, plus
%
(2) %
a compilation framework for implementing this powerful language,
in a practical way by extending a widely-used programming language, and 
levergaing best performance of logic programming systems.

We have developed such a powerful language, Alda, that combines the
advantages of logic languages and commonly-used languages for building
applications, by supporting direct use of all of logic rules, sets,
functions, updates, and objects including concurrent and distributed
processes as seamlessly integrated built-ins with no extra interfaces.
\begin{itemize}

\item Sets of rules can be specified directly
  as other definitions can,
  %
  where predicates in rules are simply
  set-valued variables holding the set of tuples for which the predicate is
  true.
  %
  Thus, predicates can be used directly as set-valued variables 
  and vice versa without needing any extra interface,
  and predicates being set-valued variables are completely different from
  functions or procedures, unlike in prior logic rule languages and
  extensions.
  %

\item 
  Queries using rule sets are calls to an inference function that computes
  desired values of derived predicates (i.e., predicates in conclusions of
  rules) given values of base predicates (i.e., predicates not in
  conclusions of rules).
  %
  Thus, queries as function calls need no extra interface either, and a
  rule set can be used with predicates in it holding the values of any
  appropriate set-valued variables.

\item Values of predicates 
  can be updated  
  either directly as for other variables or by the inference function;
  declarative semantics of rules are ensured by automatically maintaining values of derived predicates when values of base predicates are updated,
  through appropriate implicit calls to the inference function.

\item Predicates and rule sets can be object attributes 
  as well as global and local names, just as variables and functions can.

\end{itemize}
We also define\iclp{d} a formal semantics that integrates declarative and
operational semantics.  The integrated semantics supports,
seamlessly, all of logic programming with rules, database programming
with sets, functional programming, imperative programming, and
object-oriented programming including concurrent and distributed
programming. %
Note that predicates as variables,
and queries as calls with different predicate values, also
avoid the need for higher-order predicates or more sophisticated features
for reusing rules on different predicates in more complex logic languages.

Implementing such a powerful language
is nontrivial, especially to support logic rules together with updates and
objects.
We describe a compilation framework for implementation that achieves
generally good performance.

\begin{itemize}

\item The framework implements Alda by building on an object-oriented
  language that supports all other features but not logic rules, and uses
  an efficient logic rule system for queries using rules.

\item 
  The framework considers and analyzes different kinds of updates to
  predicates in different scopes
  and uses an efficient implementation for each kind to minimize calls to
  the inference function while still ensuring the declarative
  semantics of rules.

\item The framework also allows optimizations from decades of study of
  logic rules to be added for further efficiency improvements, both for
  queries using rules and for incremental queries under updates.

\end{itemize}

There has been a significant amount of related research, as discussed in
Section~\ref{sec-related}.  
Our work contains \noticlp{three}\iclp{two} main contributions:
\begin{itemize}
\item A language that supports direct use of logic rules with sets,
  functions, updates, and objects, all as built-ins, seamlessly 
  integrated\noticlp{without extra
  interfaces}\iclp{, with a formal semantics}.

\noticlp{
\item A formal semantics that integrates declarative semantics of logic
  rules with operational semantics of other constructs, both for a core
  language without objects and for a language with objects and
  classes.
}

\item A compilation framework for implementation in a widely-used
  programming language, where additional optimizations for rules can be
  exploited when available.
\end{itemize}

We have developed a prototype implementation of\iclp{ the compilation
  framework for} Alda and experimented with
a variety of programming and performance benchmarks.  Our experiments
strongly confirm the power and benefit of a seamlessly integrated language
and the generally good performance of the implementation. 
%
%
Our implementation and benchmarks are publicly available~\cite{alda23github}.

\noticlp{
The rest of the paper is organized as follows.
Section~\ref{sec-lang} presents the language.
Section~\ref{sec-impl} describes the compilation 
framework.
Section~\ref{sec-bench} presents programming and performance benchmarks.
Section~\ref{sec-expe} describes implementation and 
experimental evaluation.
Section~\ref{sec-related} discusses related work and concludes.
\noticlp{Appendix~\ref{app-formal}
defines a complete formal semantics for a language with objects and classes.
Appendix~\ref{sec-optimize} discusses related optimizations.
Appendix~\ref{app-expe} presents detailed experimental evaluation results.
}
}

\noticlp{
\mysec{Programming with rules and other features}
\label{sec-prob}

Logic rules and queries allow complex reasoning and analysis problems to be
expressed declaratively, easily and clearly, at a high level, often in ways
not possible using set expressions, recursive functions, or other
constructs.  However, logic rules are not appropriate for many other
aspects of applications, causing logic languages to include many
non-declarative features, including tricky special features
such as cut and negation as failure instead of logical
negation~\cite{Sterling:Shapiro:94}.

At the same time, languages commonly used in practice are imperative and
often object-oriented, because updates are essential in modeling the real
world,
and object encapsulation is important for organizing
system components in large applications.  In these languages, logic rules
are missing, and tedious interfaces are needed to use them,
similar to how database interfaces such as
ODBC~\cite{geiger1995inside}
and JDBC~\cite{reese2000database}
are needed to use SQL queries.
%
%

The challenge is how to support rules with all other features seamlessly,
with no extra interfaces or boilerplate code.
%

\mypar{Running example}
We use the well-known transitive closure 
problem of graphs as a running example.

Given a predicate, \co{edge}, that asserts whether there is an edge from a
first vertex to a second vertex, the transitive closure problem defines a
predicate, \co{path},
that asserts whether there is a path 
from a first vertex 
to a second vertex by following the edges.
This can be expressed in dominant logic languages such as Prolog as follows:
\begin{lstlisting}
  path(X,Y) :- edge(X,Y).
  path(X,Y) :- edge(X,Z), path(Z,Y).
\end{lstlisting}
The first rule says that 
there is a path from \co{X} to \co{Y} if there is an edge from
\co{X} to \co{Y}. The second rule says that 
there is a path from \co{X} to \co{Y} if there is an edge from \co{X} to
\co{Z} and 
there is a path from \co{Z} to \co{Y}.  Then, one can query, for example,
\begin{itemize}
  \setlength{\itemsep}{0ex}
\item[1)] the transitive closure, that is, pairs of vertices where there is
  a path from the first vertex to the second vertex, by using
  \co{path(X,Y)},
\item[2)] vertices that are reachable from a given vertex, say vertex
  \co{1}, by \co{path(1,X)},
\item[3)] vertices that can reach a given vertex, say vertex \co{2}, by
  \co{path(X,2)}, and
\item[4)] whether a given vertex, say \co{1}, can reach a given vertex, say
  \co{2}, by \co{path(1,2)}.
\end{itemize}

\mypar{Advantages of using rules over everything else for complex queries}
It is easy to see that rules are declarative and powerful, making a complex
problem easy to express, with predicates (\co{edge}, \co{path}), logic
variables (\co{X}, \co{Y}, \co{Z}), constants (\co{1}, \co{2}), and a few
symbols.  In addition, the same rules can be used easily for different
kinds of queries.
%
%
%
%

Generalizations are easy too: a predicate may have more arguments, such as
a third argument for the weight of edges; there may be more kinds of edges,
expressing more kinds of relationships; and there can be more conditions,
called hypotheses, in a rule, as well as more rules.

By using efficient inference algorithms and implementation
methods~\cite{\noticlp{LiuSto03Rules-PPDP,}LiuSto09Rules-TOPLAS}, specialization with
recursion conversion~\cite{Tek+08RulePE-AMAST}, and demand
transformation~\cite{TekLiu10RuleQuery-PPDP,TekLiu11RuleQueryBeat-SIGMOD},
queries using rules can have optimal complexities.
For example, given \m{m} edges over \m{n} vertices, the transitive closure
query can be \bigO{mn} time,
and the other three kinds of queries can be \bigO{m} time. 
With well-known logic rule engines, such as
XSB~\cite{SagSW94xsb,\noticlp{swift2012xsb,}xsb22} with its efficient emulator in C,
queries using rules can be highly efficient, even close to 
manually written C programs.

Note that rules for solving a problem may be written in different ways.
For example, for the transitive closure problem, one may reverse the order
of the rules and, in the second rule, the order of the two hypotheses or
just the two predicate names, and one may change \co{edge} to \co{path} in
the second rule.  Even though existing highly optimized rule engines may
run with drastically different complexities for these different rules, the
optimizations described
above~\cite{\noticlp{LiuSto03Rules-PPDP,}LiuSto09Rules-TOPLAS,Tek+08RulePE-AMAST,TekLiu10RuleQuery-PPDP,TekLiu11RuleQueryBeat-SIGMOD}
can give optimal complexities regardless of these different ways.

%
Without using rules, problems such as transitive closure 
could be solved by programming using imperative updates, set expressions,
recursive functions, and/or their combinations.  However, these programs are
drastically more complex or exceedingly inefficient or both.
\begin{itemize}

\item 
Using imperative updates with appropriate data structures, transitive
closure can be computed following a well-known \bigO{n^3} time algorithm,
or a better \bigO{mn} time algorithm~\cite{LiuSto09Rules-TOPLAS}.

\itempar
However, it requires using an adjacency list or adjacency matrix for graph
representation, and a depth-first search or breadth-first search for
searching the graph, updating the detailed data structures carefully as
the search progresses.  The resulting program is orders of magnitude larger
and drastically more complex.

\item
Using sets and set expressions, an imperative algorithm can be written much
more simply at a higher level.  For example, in Python, given a set \co{E}
of edges, the transitive closure \co{T} can be computed as follows, where
\co{T} starts with \co{E} (using a copy so \co{E} is not changed when
\co{T} is), workset \co{W} keeps newly discovered pairs, and \co{\WHILE
  W} continues if \co{W} is not empty:
\begin{lstlisting}
  T = E.copy()
  W = {(x,y) for (x,z) in T for (z2,y) in E if z2==z} - T
  while W:
    T.add(W.pop())
    W = {(x,y) for (x,z) in T for (z2,y) in E if z2==z} - T
\end{lstlisting}
However, expensive set operations are computed for \co{W} in each
iteration, and the total is worst-case \bigO{mn^4} time.  One may find ways
to avoid the duplicated code for computing \co{W}, but similar set
operations in each iteration cannot be avoided.
Incrementalization~\cite{PaiKoe82,Liu13book} can derive efficient \bigO{mn}
time algorithm, but such transformation is not supported in general in
any commonly-used language implementation.

\item
Using functions, one can wrap the imperative code above, either using
high-level sets or not, in a function definition that can be called at
uses.
\notes{
  In Python, this can be: \notes{trans(E) not used later anymore}
\begin{lstlisting}
  def trans(E): 
    ...  # imperative code for computing T giving E
    return T
\end{lstlisting}}

\itempar
However, if imperative updates and \co{\WHILE} loops are not allowed, the
resulting programs that use recursion would not be fundamentally easier or
simpler than the programs above.  In fact, writing these functional
programs are so nontrivial that it required research papers for individual
problems, e.g., for computing the transitive
closure~\cite{berghammer2015combining}, which develops a Haskell program
with no better than \bigO{m^3} time, which is worst-case \bigO{n^6}.

\item 
Using recursive queries in SQL, which are increasingly supported and are
essentially set expressions plus recursion, the transitive closure query
can be expressed more declaratively than the programs above.

\itempar
However, using recursive SQL queries is much more complex than using
rules~\cite{KifBL06}.
Additionally, tedious interface code is needed to use SQL
queries from a host language for programming non-SQL parts of
applications~\cite{geiger1995inside}

\end{itemize}
Moreover, all these programs are only for computing the transitive closure;
to compute the other three kinds of reachability queries, separate programs
or additional code are needed, but additional code on top of the transitive
closure code will not give the better performance possible for those
queries.

\mypar{Challenges of using rules with everything else for other tasks}
Using rules makes complex reasoning and queries easy, but other language
constructs are needed for other aspects of real-world applications.  How
can one integrate rules with everything else without extra interfaces?
There are several main challenges, for even very basic questions.

\begin{itemize}
\item The most basic question is, how do predicates and logic variables in
  logic languages relate to constructs in commonly-used languages?

  \itempar
  A well-accepted correspondence is: a predicate corresponds exactly to a
  Boolean-valued function, evaluating to true or false on given arguments.
  Indeed, in dominant logic programming languages based on
  Prolog~\cite{Sterling:Shapiro:94}, a predicate defined using rules is
  often used like a function or even procedure (when it uses non-declarative features, such as input and output) defined using expressions
  and statements in commonly-used languages.

  \itempar
  However, a big difference is that, instead of following the control flow from
  function or procedure arguments to the return value or result, techniques
  like unification are used to equate different occurrences of a variable
  in a rule.  Indeed, logic variables in rules are very different from
  variables in commonly-used languages: the former equate or relate
  arguments of predicates, whereas the latter store computed values of
  expressions.

  \itempar
  Thus, the correspondence between predicates and functions, and between
  logic variables and variables in commonly-used languages, is not proper.
  A seamlessly integrated language must establish the proper
  correspondence.

\item Another basic puzzle is even within logic languages themselves.  How
  can a set of rules defined over a particular predicate be used over a
  different predicate?  For example, how can rules defining \co{path} over
  predicate \co{edge} be used over another predicate, say \co{link}?

  \itempar 
  Supporting such uses has required higher-order
  predicates or more complex features, e.g.,~\cite{CheKW93}.  They incur
  additional baggage that compromise the ease and clarity of using rules.
  Moreover, there have not been commonly accepted constructs for them.


\item A number of other basic but important language features do not have
  commonly accepted constructs in logic languages, or are not supported at
  all: updates, classes and modules, and concurrency with threads and
  processes~\cite{maier18hist-wbook}.

  \itempar 
  Even basic declarative features are often only partially supported in
  logic languages and do not have commonly accepted constructs: 
  set comprehension, aggregation (such as count, sum, or max), and
  general quantification.

  \itempar 
  Even the semantics of recursive rules, when simple operations such as
  negation and aggregation are also used,
  has been a matter of significant
  disagreement~\cite{trusz18sem-wbook,gelfond2019vicious,LiuSto20Founded-JLC,LiuSto22RuleAgg-LFCS}.

  \itempar 
  A 
  seamlessly integrated language that supports rules must not 
  create complications for using everything else in commonly-used languages.
\end{itemize}

We address these challenges with the Alda language, with an integrated
declarative and operational semantics, allowing complex queries to be
written declaratively, easily and clearly, and be implemented with
generally good performance.

%
}

\mysec{Alda language} 
\label{sec-lang}

We first introduce rules and then describe how our overall language
supports rules with sets and functions as well as imperative updates and
object-oriented programming.  
Figure~\ref{fig-prog} shows an example program in Alda that uses all of rules, sets, functions, updates, and objects.
It will be explained throughout Sections~\ref{sec-rule}--\ref{sec-obj} when used as examples.
\noticlp{
Section~\ref{sec-formal} defines a formal semantics for a small core language
without objects.
}%
A complete exposition of the\noticlp{ abstract syntax and} formal semantics is in\iclparxiv{~\ref{app-formal}}{~\cite[Appendix]{Liu+23RuleLangInteg-arxiv}}\oops{
the Supplementary Material}.

\begin{figure}[t]
\centering
\begin{lstlisting}[numbers=left, basicstyle=\scriptsize\ttfamily]
class CoreRBAC:                  # class for Core RBAC component/object
  def setup():                   # method to set up the object, with no arguments
    self.USERS, self.ROLES, self.UR := {},{},{}
                                 # set users, roles, user-role pairs to empty sets
  def AddRole(role):             # method to add a role
    ROLES.add(role)              # add the role to ROLES
  def AssignedUsers(role):       # method to return assigned users of a role
    return {u: u in USERS | (u,role) in UR}  # return set of users having the role
\end{lstlisting}
%
%
\Vex{-2}\iclp{\Vex{-1}}
\begin{lstlisting}[basicstyle=\scriptsize\ttfamily]
  ...
\end{lstlisting}
\Vex{-2}
\begin{lstlisting}[numbers=left,firstnumber=9, basicstyle=\scriptsize\ttfamily]
class HierRBAC extends CoreRBAC: # Hierarchical RBAC extending Core RBAC
  def setup():
    super().setup()              # call setup of CoreRBAC, to set sets as in there
    self.RH := {}                # set ascendant-descendant role pairs to empty set
  def AddInheritance(a,d):       # to add inherit. of an ascendant by a descendant
    RH.add((a,d))                # add pair (a,d) to RH
  rules trans_rs:                # rule set defining transitive closure
    path(x,y) if edge(x,y)       # path holds for (x,y) if edge holds for (x,y)
    path(x,y) if edge(x,z), path(z,y)  # ... if edge(x,z) holds and path(z,y) holds
  def transRH():                 # to return transitive RH and reflexive role pairs
    return infer(path, edge=RH, rules=trans_rs) + {(r,r): r in ROLES}
  def AuthorizedUsers(role):     # to return users having a role transitively
    return {u: u in USERS, r in ROLES | (u,r) in UR and (r,role) in transRH()}
\end{lstlisting}
%
\Vex{-2}\iclp{\Vex{-1}}
\begin{lstlisting}[basicstyle=\scriptsize\ttfamily]
  ...
\end{lstlisting}
\Vex{-2}
\begin{lstlisting}[numbers=left,firstnumber=22, basicstyle=\scriptsize\ttfamily]
h = new(HierRBAC, [])            # create HierRBAC object h, with no args to setup
h.AddRole('chair')               # call AddRole of h with role 'chair'
\end{lstlisting}
\Vex{-2}\iclp{\Vex{-1}}
\begin{lstlisting}[basicstyle=\scriptsize\ttfamily]
...
\end{lstlisting}
\Vex{-2}
\begin{lstlisting}[numbers=left,firstnumber=24, basicstyle=\scriptsize\ttfamily]
h.AuthorizedUsers('chair')       # call AuthorizedUsers of h with role `chair'
\end{lstlisting}
\Vex{-2}\iclp{\Vex{-1}}
\begin{lstlisting}[basicstyle=\scriptsize\ttfamily]
...
\end{lstlisting}\noticlp{\iclp{\Vex{-.5}}}
\Vex{-2}
\caption{An example program in Alda, for Role-Based Access Control (RBAC), demonstrating logic rules used with sets, functions, updates, and objects.\iclp{\Vex{-2}}}
\label{fig-prog}
\end{figure}

\mysubsec{Logic rules}
\label{sec-rule}

We support rule sets of the following form, where
\p{name} is the name of the rule set, \p{declarations} is a set of
predicate declarations, and the body is a set of rules.\iclp{$\!\!$}
\begin{code}
  \RULES \p{name} (\p{declarations}): 
    \p{rule}+
\end{code}
A \defn{rule} is either one of the two equivalent forms below (for users
accustomed to either form), meaning
that if \p{hypothesis\sb{1}} through \p{hypothesis\sb{h}} all hold, then
\p{conclusion} holds.
\begin{code}
  \p{conclusion} \cIF \p{hypothesis\sb{1}}, \p{hypothesis\sb{2}}, \p{...}, \p{hypothesis\sb{h}}
  \cIF \p{hypothesis\sb{1}}, \p{hypothesis\sb{2}}, \p{...}, \p{hypothesis\sb{h}}: \p{conclusion}
\end{code}
If a conclusion holds without a hypothesis, 
then \co{if} and \co{:} are omitted.

Declarations are about predicates used in the rule set, for advanced uses, and are optional.
For example, they may specify 
argument types of predicates, so rules can be compiled to efficient
standalone imperative
programs~\cite{\noticlp{LiuSto03Rules-PPDP,}LiuSto09Rules-TOPLAS} that are expressed
in typed languages~\cite{RotLiu07Retrieval-PEPM}.
They may also specify assumptions about predicates~\cite{LiuSto20Founded-JLC}
to support different desired semantics~\cite{LiuSto21LogicalConstraints-JLC,LiuSto22RuleAgg-JLC}.
We omit the details because they are orthogonal to the focus of the
paper.  In particular, we omit types to avoid unnecessary clutter in code.

%
We use Datalog rules~\cite{AbiHulVia95,maier18hist-wbook} in examples,
but our method of integrating semantics applies to rules in general. 
Each hypothesis and conclusion in a rule is an \defn{assertion},
of the form\Vex{-.5}\iclp{\Vex{\noticlp{-2}\iclp{-1.2}}}
\noticlp{\[}\iclp{\begin{center}\Vex{0}}
\co{\p{p}(\p{arg\sb{1}},\p{...},\p{arg\sb{a}})}
\iclp{\end{center}}\noticlp{\]}\iclp{\Vex{\noticlp{-2.2}\iclp{-1.2}}}
where \p{p} is a \defn{predicate}, and each \p{arg_k} is a variable 
or a constant.
We use numbers and quoted strings to represent constants, and the rest are
variables.
As is standard for safe rules, all variables in the conclusion must be in a hypothesis.
If a conclusion holds without a hypothesis, 
then each argument in the conclusion must be a constant, in which case the
conclusion is called a \defn{fact}.
Note that a predicate is also called a \defn{relation}, relating the
arguments of the predicate.
%

\begin{example}
  For computing the transitive closure of a graph in the running example,
  the rule set, named \co{trans\_rs}, in Figure~\ref{fig-prog} (lines 15-17) can be written.
%
%
  The rules are the same as in dominant logic languages except for
  the use of lower-case variable names, the change of \co{:-} to \co{if}, and
  the omission of dot at the end of each rule.
  %
\end{example}

\mypar{Terminology}
Consider a set of rules.  Predicates not in any conclusion are called \defn{base
  predicates}, and the other predicates are called
\defn{derived predicates}.
%
We say that a predicate \p{p} \defn{depends on} a predicate \p{q} if \p{p}
is in the conclusion of a rule whose hypotheses contain \p{q} or contain a
predicate that depends on \p{q} recursively.
We say that a derived predicate \p{p} \defn{fully
  depends on} a set \p{s} of base predicates if \p{p} does not
depend on other base predicates.

\begin{example}
  In rule set \co{trans\_rs}, \co{edge} is a base predicate, and
  \co{path} is a derived predicate.  \co{path} depends on \co{edge} and
  itself.  \co{path} fully depends on \co{edge}.
\end{example}

\mysubsec{Integrating rules with sets, functions, updates, and objects}
\label{sec-integrate}


Our overall 
language supports all of rule sets and the following language constructs as
built-ins; all of them can appear in any scope---global, class, and local.
\begin{itemize}

\item Sets and set expressions (comprehension, aggregation, quantification,
  and high-level operations such as union) to make non-recursive queries
  over sets easy to express.

\item Function and procedure definitions with optional keyword arguments,
  and function and procedure calls.

\item Imperative updates by assignments and membership changes, to sets and
  data of other types, in sequencing, branching, and looping statements.

\item Class definitions containing object field and method (function and procedure) definitions, object creations, and inheritance.

\end{itemize}
A name holding any value is \defn{global} if it is introduced (declared or
defined) 
at the global scope; 
is an \defn{object field} if it is introduced
for that object;
or is \defn{local} to the function, method, or rule set that contains it
otherwise.
After a name is defined, the value that it is holding is available:
globally for a global name, on the object for an object field, and in the
enclosing function, method, or rule set for a local name.
\notes{
A rule set can also be in any scope as other built-ins, and usual name
lookup is used to locate a rule set using its name.  Predicate names are
also treated as other names.}

\begin{example}
  Rule set \co{trans\_rs} in Figure~\ref{fig-prog} (defined on lines 15-17 and queried using a call to an inference function. \co{infer}, on line 19) is used together with sets (defined on lines 3 and 12), set expressions (on lines 8, 19, and 21), functions (defined on lines 7-9, 18-19, and 20-21), procedures (defined on lines 2-3, 5-6, 10-12, and 13-14), updates (on lines 3, 6, 12, 14), classes (defined on lines 1 and 9, with inheritance), and objects (created on line 22).
  No extra code is needed to convert \co{edge} and \co{path}, declare logic variables, and so on.
\end{example}

The key ideas of our seamless integration of rules with sets, functions,
updates, and objects are:
(1) a predicate is a set-valued variable that holds the set of tuples for
which the predicate is true,
(2) queries using rules are calls to an inference function that computes
desired sets using given sets,
%
(3) values of predicates can be updated either directly as for other
variables or by the inference function, and
%
(4) predicates and rule sets can be object attributes 
as well as global and local names, just as sets and functions can.

\mypar{Integrated semantics, ensuring declarative semantics of rules}
In our overall language, the meaning of a rule set \p{rs} is completely
declarative, exactly following the standard least fixed-point semantics of
rules~\cite{fitting2002fixpoint,LiuSto09Rules-TOPLAS}:
\begin{itemize}
\item[] Given values of any set \p{s} of base predicates in \p{rs}, the
  meaning of \p{rs} is, for all derived predicates in \p{rs} that fully
  depend on \p{s}, the least set of values that can be inferred, directly
  or indirectly, by using the given values and the rules in \p{rs};

  for any derived predicate in \p{rs} that does not fully depend on
  \p{s},
  i.e., depends on any base predicate whose values are not given, its value
  is \defn{undefined}.
\end{itemize}
The operational semantics for the rest of the language ensures this
declarative semantics of rules.  
The precise constructs for using rules with sets, functions, updates, and
objects are described in Sections~\ref{sec-pred}--\ref{sec-obj}.



\mysubsec{Predicates as set-valued variables}
\label{sec-pred}
For rules to be easily used
with everything else, our most basic principle in 
designing the language
is to treat a predicate as a set-valued variable that holds the set of tuples
that are true for the predicate, that is:\iclp{\Vex{-1}}
\begin{myquote}{3ex} 
  For any predicate \co{\p{p}} over values \co{\p{x\sb{1}},\p{...},\p{x\sb{a}}},
  assertion \co{\p{p}(\p{x\sb{1}},\p{...},\p{x\sb{a}})} is true---i.e.,
  \co{\p{p}(\p{x\sb{1}},\p{...},\p{x\sb{a}})} is a fact---if and only if
  tuple \co{(\p{x\sb{1}},\p{...},\p{x\sb{a}})} is in set \co{\p{p}}.
  Formally,\Vex{-1}\noticlp{\iclp{\Vex{-1}}}\iclp{\Vex{-.5}}
\noticlp{\[}\iclp{\begin{center}$\Vex{-1}}
\co{\p{p}(\p{x\sb{1}},\p{...},\p{x\sb{a}})} 
~\Longleftrightarrow~ 
\co{(\p{x\sb{1}},\p{...},\p{x\sb{a}}) \IN \p{p}}
\iclp{$\end{center}}\noticlp{\]}
\end{myquote}
This means that, as variables,
predicates in a rule set can be introduced in any scope---as 
global variables, 
object fields, 
or variables local to the rule set---and 
they can be written into and read from without needing any extra
interface.

\begin{example}
  In rule set \co{trans\_rs} in Figure~\ref{fig-prog}, predicate \co{edge}
  is exactly a variable holding a set of pairs, such that
  \co{edge(\m{x},\m{y})} is true iff \co{(\m{x},\m{y})} is in \co{edge}, and \co{edge} is local to \co{trans\_rs}. 
  In general, \co{edge} can be a global variable,
  an object field, or a local variable of \co{trans\_rs}.
  Similarly for predicate \co{path}.
\end{example}

Writing to predicates is discussed later under updates to predicates, but
reading and using values of predicates can simply use all operations on
sets.
We use set expressions including the following:
\Vex{1}\\
{\noticlp{\iclp{\scriptsize}}\iclp{\footnotesize}
\begin{tabular}{@{\Hex{\noticlp{2}\iclp{3.5}}}l@{\Hex{1}\Hex{10}}l}
    \co{\p{exp} \IN \p{sexp}}
    & \Hex{-0}membership\\
    \co{\p{exp} \NOT \IN \p{sexp}}
    & \Hex{-0}negated membership\\
    \co{\p{sexp\sb{1}} + \p{sexp\sb{2}}}
    & \Hex{-0}union\\
    \co{\{\p{exp}:\,\p{v\sb{1}} \IN \p{sexp\sb{1}},\p{...},\p{v\sb{k}} \IN \p{sexp\sb{k}}\,|\,\p{bexp}\}}
    & comprehension\\
    \co{\p{agg} \p{sexp}}, \Hex{1} where \co{\p{agg}} is \co{count}, \co{max}, \co{min}, \co{sum}
    & aggregation \\
    \co{\SOME~\p{v\sb{1}} \IN \p{sexp\sb{1}},\p{...},\p{v\sb{k}} \IN \p{sexp\sb{k}}\,|\,\p{bexp}}
    & existential quantification
\end{tabular}
}\Vex{1}\\
A comprehension returns the set of values of 
\p{exp} for all combinations of values of variables that satisfy
all membership clauses \co{\p{v_i} \IN \p{sexp_i}} and condition 
\p{bexp}.
%
An aggregation returns the count, max, etc. of the set value of \p{sexp}.
An existential quantification returns true iff for some 
combination of values of variables that satisfies all \co{\p{v_i} \IN \p{sexp}} clauses, condition \p{bexp} holds. When an existential quantification returns true, variables \p{v_1},...,\p{v_k} are bound to a witness.
Note that these set queries, as in~\cite{Liu+17DistPL-TOPLAS}, are more powerful than those in Python.

\begin{example}
  For computing the transitive closure \co{T} of a set \co{E} of edges, the
  following \co{\WHILE} loop with quantification can be used (we will see
  that we use objects and updates as in Python except for  the syntax \co{:=}
  for assignment in this paper):
\begin{lstlisting}
  T := E.copy()
  while some (x,z) in T, (z,y) in E | (x,y) not in T:
    T.add((x,y))
\end{lstlisting}\iclp{\Vex{-3}}\noticlp{
This is simpler than the Python \co{\WHILE} loop in Section~\ref{sec-prob}:
it finds a witness containing \co{x} and \co{y} directly using \co{some}, 
instead of constructing workset \co{W} and then using \co{pop} to get a pair.}
\end{example}

In the comprehension and aggregation forms, each \p{v_i} can also be a
tuple pattern that elements of the set value of \p{sexp_i} must
match~\cite{Liu+17DistPL-TOPLAS}.
A \defn{tuple pattern} is 
a tuple in which each component is a non-variable expression,
a variable possibly prefixed with \co{=}, a wildcard \co{\_}, or
recursively a tuple pattern.
For a value to match a tuple pattern, 
it must have the corresponding tuple structure, with corresponding
components equal the values of non-variable expressions and variables
prefixed with \co{=}, and with corresponding components assigned to
variables not prefixed with \co{=}; multiple occurrences of a variable must
be assigned the same value; corresponding components of wildcard are
ignored. 

\begin{example}
  To return the set of second component of pairs in \co{path} whose first
  component equals the value of variable \co{x}, and where that second
  component is also the first component of pairs in \co{edge} whose second
  component is 1, one may use a set comprehension with tuple patterns:
\begin{lstlisting}
  {y: (=x,y) in path, (y,1) in edge}
\end{lstlisting}\Vex{-3}
\end{example}

Now that predicates in rules correspond to set-valued variables, instead of
functions or procedures, we can further see that logic variables, i.e.,
variables in arguments of predicates in rules,
are like pattern variables, i.e., variables not
prefixed with \co{=} in patterns.
%
These variables are used for relating values, through what is generally
called unification;
they do not hold values, unlike 
variables prefixed with \co{=} in patterns.

\mysubsec{Queries as calls to an inference function} 
\label{sec-infer}
For inference and queries using rules, calls to a built-in inference
function \co{\INFER}, 
of the following form, are used, with
\p{query\sb{k}}'s and \co{\p{p\sb{k}}=\p{sexp\sb{k}}}'s being optional:\m{\!}
%
\begin{code}
  \INFER(\p{query\sb{1}},\,\p{...},\,\p{query\sb{j}}, \p{p\sb{1}}=\p{sexp\sb{1}},\,\p{...},\,\p{p\sb{i}}=\p{sexp\sb{i}}, \RULES=\p{rs})
\end{code}
\co{\p{rs}} is the name of a rule set.  Each \co{\p{sexp\sb{k}}} is a
set-valued expression.  Each \co{\p{p\sb{k}}} is a base predicate of \p{rs}
and is local to \p{rs}.
Each \co{\p{query\sb{k}}} is of the form
\co{\p{p}(\p{arg_1},\p{...},\p{arg_a})},
where \co{\p{p}} is a derived predicate of \p{rs}, and each 
argument \co{\p{arg_k}} is a constant, 
a variable possibly prefixed with
\co{=}, or wildcard \co{\_}.
A variable prefixed with \co{=} indicates a bound variable whose value will
be used as a constant when evaluating the query.
So arguments of queries are patterns too.
If all \co{\p{arg_k}}'s are \co{\_}, the abbreviated form \co{\p{p}} 
can be used.

Function \co{\INFER} can be called implicitly by the language implementation
or explicitly by the user.
%
It is called automatically as needed and can be called explicitly when desired.

\begin{example} 
  For inference using rule set \co{trans\_rs} in Figure~\ref{fig-prog}, where
  \co{edge} and \co{path} are local variables, \co{\INFER} can be called in
  many ways, including: 
\begin{lstlisting}
  infer(path, edge=RH, rules=trans_rs)
  infer(path(_,_), edge=RH, rules=trans_rs)
  infer(path(1,_), path(_,=R), edge=RH, rules=trans_rs)
\end{lstlisting} 
The first is as in Figure~\ref{fig-prog} (line 19).
The first two calls are equivalent: \co{path} and \co{path(\_,\_)} both
query the set of pairs of vertices having a path from the first vertex to
the second vertex, 
following edges given by the value of variable \co{RH}.
In the third call, \co{path(1,\_)} queries the set of vertices having a
path from vertex 1, and \co{path(\_,=R)} queries the set of vertices having
a path to the vertex that is the value of variable \co{R}.

If \co{edge} or \co{path} is a global variable or an object field, 
one may call \co{\INFER} on \co{trans\_rs} without 
assigning to \co{edge} or querying 
\co{path}, respectively.
\end{example}

The operational semantics of a call to \co{\INFER}
is exactly like other function calls, except for the special forms of
arguments and return values, and of course the inference function performed
inside:
\begin{enumerate}

\item[1)] For each value \p{k} from 1 to \p{i}, assign the set value of
  expression \p{sexp_k} to predicate \p{p_k} that is a base predicate of
  rule set \p{rs}.

\item[2)] Perform inference using the rules in \p{rs} and the given values
  of base predicates of \p{rs} following the declarative semantics,
  including assigning to derived predicates that are not local.

\item[3)] For each value \p{k} from 1 to \p{j},
return the result of query \p{query_k} as the \p{k}th component of the
return value.
The result of a query with \m{l} distinct variables not prefixed with \co{=} is a set of tuples of
\m{l} components, one for each of the distinct variables in their order of
first occurrence in the query.
\end{enumerate}
Note that
when there are no \co{\p{p\sb{k}}=\p{sexp\sb{k}}}'s, only defined values of
base predicates that are not local to \p{rs} are used; and
when there are no \p{query\sb{k}}'s, only values of derived predicates that
are not local to \p{rs} may be inferred and no value is returned.
This is the case for implicit calls to \co{\INFER} on \p{rs}.

\noticlp{
Section~\ref{sec-rbac} 
on benchmarks using Role-Based Access Control (RBAC) discusses different ways
of using rules and different
calls to \co{\INFER}: implicit vs.\ explicit, in an enclosing expression
vs.\ by itself, passing in all base predicates vs.\ only some, etc. 
}

%

\mysubsec{Updates to predicates}
\label{sec-upd}
%
%
Values of base predicates can be updated directly as for other set-valued
variables, and values of derived predicates are updated by the inference
function.

Base predicates of a rule set \p{rs} that are local to \p{rs} are assigned
values at calls to \co{\INFER} on \p{rs}, as described earlier.  Base
predicates that are not local can be updated by\noticlp{ using} assignment statements
or set\noticlp{ membership} update operations.  
We use
\begin{code}
  \p{lexp} := \p{exp}
\end{code} 
for assignments, where \co{\p{lexp}} can
also be a nested tuple of variables, and each variable
is assigned the corresponding component of the value of \co{\p{exp}}. 

Derived predicates of a rule set \p{rs} can be updated only by calls to the
inference function on \p{rs}.  The updates must ensure the declarative
semantics of~\p{rs}:
\begin{itemize}

\item[] Whenever a base predicate of \p{rs} is updated in the program, the
  values of the derived predicates in \p{rs} are maintained
  according to the declarative semantics of \p{rs} by
  calling \co{\INFER} on \p{rs}.

  Updates to derived predicates of \p{rs} outside \p{rs} are not allowed,
  and any violation will be detected and reported at compile time if
  possible and at runtime otherwise.


\end{itemize}
Simply put, updates to base predicates trigger updates to derived
predicates, and other updates to derived predicates are not allowed.
This ensures the invariants that the derived predicates hold the values
defined by the rule set based on values of the base
predicates, as required by the declarative semantics.
Note that this is the most straightforward semantics, but the
implementation can avoid many inefficiencies with optimizations\noticlp{,
  as described in Section~\ref{sec-optimize}}.

\begin{example}
  Consider rule set \co{trans\_rs} in Figure~\ref{fig-prog}. 
  If \co{edge} is not local, one may assign a set of pairs to \co{edge}:
\begin{lstlisting}
  edge := {(1,8),(2,9),(1,2)}
\end{lstlisting}\iclp{\Vex{-1}}

If \co{edge} is local, the calls to \co{\INFER} in the example in
Section~\ref{sec-infer} assign the value of \co{RH} to \co{edge}.

If \co{path} is not local, then a call \co{\INFER(edge=RH, \RULES=trans\_rs)}
updates \co{path}, contrasting the first two calls to \co{\INFER} in the
example in Section~\ref{sec-infer} that return the value of \co{path}.

If \co{path} is local, the return value of \co{\INFER} can be assigned to
variables.  For example, for the third call to \co{\INFER} in the example in
Section~\ref{sec-infer}, this can be
\begin{lstlisting}
  from1,toR := infer(path(1,_), path(_,=R), edge=RH, rules=trans_rs)
\end{lstlisting}\iclp{\Vex{-.5}}

If both \co{edge} and \co{path} are not local, then whenever \co{edge} is
updated,
an implicit call\oops{\linebreak} \co{\INFER(\RULES=trans\_rs)} is made automatically to
update \noticlp{the value of }\co{path}.
\end{example}

\iclp{
For the RBAC example in Figure~\ref{fig-prog}, different ways of using rules are possible, including (1) allloc: adding a rule \co{path(x,x) if role(x,x)} to the rule set, adding \co{role=ROLES} in the call to \co{infer}, and removing the union in function \co{transRH}, so all predicates are local variables; (2) nonloc: as in allloc, except to replace predicates \co{edge}, \co{role}, and \co{path} with \co{RH}, \co{ROLES}, and a new field \co{transRH}, respectively, replace call \co{transRH()} with field \co{transRH}, and remove function \co{transRH}; (3) union: as in Figure~\ref{fig-prog}; and other combinations of aspects of (1)--(3).
}

\mysubsec{\oops{\Hex{0}}Using predicates and rules with objects and classes} 
\label{sec-obj}
Predicates and rule sets can be object fields as well as global and local
names, just as sets and functions can, as discussed in
Section~\ref{sec-integrate}.  This allows predicates and rule sets to be
used seamlessly with objects
in object-oriented programming. 


%
For other constructs than those described above, we use those in high-level
object-oriented languages.  We mostly use Python syntax (looping,
branching, indentation for scoping, `\co{:}' for elaboration, `\co{\#}' for
comments, etc.) for succinctness, but with a few conventions from Java
(keyword \co{new} for object creation, 
keyword \co{extends} for subclassing,
and omission of \co{self}, the equivalent of \co{this} in Java, 
when there is no ambiguity)
for ease of reading.

\begin{example}
We use 
Role-Based Access Control (RBAC) to show the need of using rules with all
of sets, functions, updates, and objects and classes.

RBAC is a security policy framework for controlling user access to
resources based on roles 
and is widely used in large organizations.  The ANSI standard for
RBAC~\cite{ansi04role} was approved in 2004 after several 
rounds of public
review~\cite{Sandhu+00,Jaeger:Tidswell:00,ferraiolo01proposed}, building on
much research during the preceding decade and earlier.
High-level executable specifications were developed for the entire RBAC
standard~\cite{LiuSto07RBAC-ONR}, where all queries are declarative except
for computing the transitive role-hierarchy relation in Hierarchical RBAC,
which extends Core RBAC.

Core RBAC defines functionalities relating users, roles, permissions, and
sessions. 
It includes the sets and update and query functions
in class \co{CoreRBAC} in Figure~\ref{fig-prog}, as in~\cite{LiuSto07RBAC-ONR}.\footnote{Only a few selected sets and
  functions are included, and with small changes to names and syntax.}

Hierarchical RBAC adds support for a role hierarchy, \co{RH}, and update
and query functions extended for \co{RH}.  It includes the update and query
functions in class \co{HierRBAC} in Figure~\ref{fig-prog}, as in~\cite{LiuSto07RBAC-ONR},\m{^1} except that function \co{transRH()} in~\cite{LiuSto07RBAC-ONR} computes the
transitive closure of \co{RH} plus reflexive role pairs for all roles in
\co{ROLES} by using a complex and inefficient \co{\WHILE} loop
\noticlp{similar to that 
in Section~\ref{sec-prob}}\iclp{much worse than that in Section~\ref{sec-pred} (due to Python's lack of \co{some} with witness)}
plus a union with the set of reflexive role pairs
\co{\{(r,r):~r in ROLES\}}, whereas function \co{transRH()} in Figure~\ref{fig-prog} simply calls \co{\INFER} and unions the result with
reflexive role pairs.
%
%
%

Note though, in the RBAC standard, a relation \co{transRH} is used in place
of \co{transRH()}, intending to maintain the transitive role hierarchy
incrementally while \co{RH} and \co{ROLES} change.  It is believed that this is done
for efficiency, because the result of \co{transRH()} is used continually,
while \co{RH} and \co{ROLES} change infrequently.  However, the maintenance
was done inappropriately~\cite{LiuSto07RBAC-ONR,Li+07critique} and
warranted the use of \co{transRH()} to ensure correctness before
efficiency.

%
%

%

Overall, the RBAC specification relies extensively on all of updates, sets,
functions, and objects and classes with inheritance, besides rules:
(1) updates for setting up and updating the state of the RBAC system,
(2) sets and set expressions for holding the system state and expressing
set queries exactly as specified in the RBAC standard,
(3) methods and functions for defining and invoking update and query
operations\noticlp{, including
function \co{transRH()}}, and
(4) objects and classes for capturing different components---\co{CoreRBAC}, \co{HierRBAC}, constraint RBAC, their further refinement, extensions, and
combinations, totaling 9 components, corresponding to 9 classes, including
5 subclasses of \co{HierRBAC}~\cite{ansi04role,LiuSto07RBAC-ONR}.
\end{example}

\noticlp{
\input{semantics-core}
}

\mysec{Compilation\noticlp{ and optimization}}
\label{sec-impl}

We describe our compilation framework for implementing Alda, by building on
an object-oriented language that supports all features except rules and
queries and on an efficient logic rule engine for queries using
rules. Three main tasks are (1) compiling rule sets
to generate rules accepted by the rule engine, (2) compiling queries using
rules to generate queries accepted by the rule engine, together with
automatic conversion of data and query results, and (3) compiling updates
to predicates that require implicit automatic queries and updates of the
query results.\noticlp{

}
The compiler must appropriately handle scoping of rule sets and
predicates for all three tasks.  Besides that, task (1) is
straightforward, task (2) is also straightforward but tedious, and task (3) requires the most
analysis, so we focus on task (3) below.

We first describe how to compile all possible updates to predicates,
starting with the checks and actions needed to correctly handle updates
for a single rule set with implicit and explicit calls to \co{\INFER}.
We then describe how to implement the inference in \co{\INFER}.
In\iclp{~\ref{sec-optimize}}\noticlp{ the last
  subsection}\oops{ the Supplementary Material}, we systematize powerful
optimizations that can be added in the overall compilation framework;
clearly separated handling of updates and queries in our compilation
framework allows optimizations to be added in a modular fashion.

\mysubsec{Compiling updates to predicates}

The operational semantics to ensure the declarative semantics of a rule set \p{rs} is
conceptually simple, but for efficiency, the implementation required
varies, depending on the kind of updates to base predicates of \p{rs}
outside \p{rs}.  
Note that inside \p{rs} there are no updates to base predicates of
\p{rs}, by definition of base predicate.

\begin{enumerate}
\item[1)] {\bf Local updates.} Local variables of \p{rs}, i.e., predicates local to \p{rs},
  can be assigned values only at explicit calls to \co{\INFER} on \p{rs}.
  Such a call passes in values of local variables that are base predicates
  of \p{rs} before doing the inference.
  Values of local variables that are derived predicates of \p{rs} can only
  be used in constructing answers to the queries in the call, and the
  answers are returned from the call.

  \itempar
  There are no updates outside \p{rs} to local variables that are 
  derived predicates of \p{rs}, 
  by definition of local variables.

\item[2)] {\bf Non-local updates.} For updates to non-local
  variables of \p{rs},
  an implicit call to \co{\INFER} on \p{rs} needs to be made only after
  every update to a base predicate of \p{rs}.
  

  \itempar

  Statements outside \p{rs} that update derived
  predicates of \p{rs} are identified and reported as errors. 

  \itempar 

  In languages or application programs where variables hold data values,
  such as in database languages and applications, these updates can be
  determined simply at compile time, e.g., if \co{s} holds a set value,
  then \co{s\,:=\,s+\{x\}} updates the set value of \co{s}.  This is also
  the case when logic rules are used in these languages and programs.

  \itempar 

  In programs where variables may be references to data values, 
  each update needs to check whether the updated variable may alias a
  predicate of \p{rs}, conservatively at compile-time if possible, and at
  runtime otherwise.




\end{enumerate}
%
%
%
%
%
%
%
To satisfy these requirements, the overall method for compiling an
update to a variable \co{v} outside rule sets 
is:
\begin{itemize}

\item 
  In languages or application programs where variables hold data values,
  report a compile-time error if \co{v} is a derived predicate of any rule
  set; otherwise, for each rule set \p{rs} that contains \co{v} as a base
  predicate, insert code, after the update,
  that calls \co{\INFER} on \p{rs} with no arguments for base predicates and no
  queries.

\item 
  Otherwise, if \co{v} may refer to a predicate in a rule set,
  insert code that does the following after the update:
  if \co{v} refers to a derived predicate of any rule set, report a runtime
  error and exit; otherwise for each rule set \p{rs}, if \co{v} refers to
  a base predicate of \p{rs}, call \co{\INFER} on \p{rs} with no arguments
  for base predicates and no queries.

\end{itemize}
Our method for compiling an explicit call to \co{\INFER} on a rule set
directly follows the operational semantics of \co{\INFER}.

In effect, function \co{\INFER} 
is called to implement a wide range of control: from inferring everything
possible using all rule sets and values of all base predicates at every
update\noticlp{ in the most extensive case}, to answering specific queries using specific rules and specific
sets of values of specific base predicates at explicit calls.

Obviously, updates 
in different cases may have significant impact on program efficiency.
Update analysis is needed to determine the case and generate correct code\noticlp{ in all cases}.
Our compilation method above minimizes calls to \co{\INFER} in each case. 


\Vex{-.5}
\mysubsec{Implementing inference and queries} 

Any existing method can be used to implement the functionality inside
\co{\INFER}.  The inference and queries for a rule set can use either
bottom-up or top-down
evaluation~\cite{KifLiu18wbook,TekLiu10RuleQuery-PPDP,TekLiu11RuleQueryBeat-SIGMOD},
so long as they use the rule set and values of the base
predicates according to the declarative semantics of rules

The inference and queries can be either performed by using a general logic
rule engine, e.g., XSB~\cite{SagSW94xsb,\noticlp{swift2012xsb,}xsb22}, 
or compiled to specialized standalone executable code as in,
e.g.,~\cite{LiuSto09Rules-TOPLAS,RotLiu07Retrieval-PEPM,jordan2016souffle},
that is then executed.
Our current implementation uses the former approach, by indeed using the
well-known XSB system, as described in Section~\ref{sec-expe}, because it
allows easier extensions to support more kinds of rules and optimizations
that are already supported in XSB.
Other powerful logic rule engines, including efficient Answer Set
Programming (ASP) systems such as Clingo~\cite{gebser2019multi}, can
certainly be used also.


\noticlp{
\oops{\pagebreak}\section{Powerful optimizations}
\label{sec-optimize}
Efficient inference and queries using rules is well known to be challenging
in general, and especially so if it is done repeatedly to ensure the
declarative semantics of rules under updates to predicates.  Addressing the
challenges has produced an extensive literature in several main areas in
computer science---database, logic programming, automated reasoning, and
artificial intelligence in general---and is not the topic of this paper.

Here, we describe how well-known analyses and optimizations can be used
together to improve the implementation of the overall language as well as
the rule language, giving a systemic perspective of all main optimizations
for efficient implementations.  There are two main areas of optimizations.

The first area is for inference under updates to the predicates used.
There are three main kinds of optimizations in this area: (1) reducing
inference triggered by updates, (2) performing inference lazily only when
the results are demanded, and (3) doing inference incrementally when
updates must be handled to give results:

\begin{description}
\item[Reducing update checks and inference.]
In the presence of aliasing, it can be extremely inefficient to check, for
all rule sets after every update, that the update is not to a derived
predicate of the rule set and whether a call to \co{\INFER} on the rule set
is needed, not knowing statically whether the update affects a base
predicate of the rule set.
Alias analysis, e.g.,~\cite{Goy05,Gor+10Alias-DLS}, can help reduce such
checks by statically determining updates to variables that possibly alias a
predicate of a rule set.

\item[Demand-driven inference.] 
%
Calling \co{\INFER} after every update to a base predicate can be inefficient and wasteful,
because updates can occur frequently while the maintained derived
predicates are rarely used.
To avoid this inefficiency, \co{\INFER} can be called on demand just before
a derived predicate is used, e.g.,~\cite{FonUll76,RotLiu08OSQ-GPCE,Liu+16IncOQ-PPDP},
instead of immediately after updates to base predicates.

\item[Incremental inference.] 
More fundamentally, even when derived predicates are frequently used,
\co{\INFER} may be called repeatedly on slightly changed or even
unchanged base predicates, in which case computing the results from scratch is extremely wasteful.
Incremental computation can drastically reduce this inefficiency by
maintaining the values of derived predicates
incrementally, e.g.,~\cite{GupMum99maint,SahaRam03}.

\end{description}

The second area is for efficient implementation of  rules by themselves,
without considering updates to the predicates used.  There are two main
groups of optimizations.

\begin{description}
\item[Internal demand-driven and incremental inference.] 
Even in a single call to \co{\INFER}, significant
optimizations are needed. 

In top-down evaluation (which is already driven by the given query as
demand), subqueries can be evaluated repeatedly, so
tabling~\cite{TamSat86,CheWar96} (a special kind of incremental computation
by memoization) is critical for avoiding not only repeated evaluation of
queries but also non-termination when there is recursion.

In bottom-up evaluation (which is already incremental from the ground up),
demand
transformation~\cite{TekLiu10RuleQuery-PPDP,TekLiu11RuleQueryBeat-SIGMOD},
which improves over magic sets~\cite{Ban+86,AbiHulVia95} exponentially, can transform
rules to help avoid computations not needed to answer the given query.

\item[Ordering and indexing for inference.]
Other factors can also drastically affect the performance of logic queries
in a single call to \co{\INFER}~\cite{maier18hist-wbook,liu18LPappl-wbook}.  

Most prominently, in dominant logic rule engines like XSB, changing the
order of joining hypotheses in a rule can impact performance dramatically,
e.g., for the transitive closure example, reversing the two hypotheses in
the recursive rule can cause a linear factor performance difference.
Reordering and indexing~\cite{LiuSto09Rules-TOPLAS,Liu+16IncOQ-PPDP}
are needed to avoid such severe slowdowns.
\end{description}

}

\noticlp{
\input{bench}
}


\notes{
We also extend the language to support nested scopes, where
classes and functions can be arbitrarily nested, and rule sets can be in
any nested scope.
Name keeping and lookup in nested scopes use standard methods as for other names. 
Other than that, our implementation method applies as described, because it
only differentiates local vs non-local updates.
} 

\mysec{Implementation and experimental evaluation}
\label{sec-expe}

We have 
implemented a prototype compiler for Alda.
%
The compiler generates executable code in Python.  The generated code calls the XSB
logic rule engine~\cite{SagSW94xsb,\noticlp{swift2012xsb,}xsb22} for inference using rules.


%
%

We implemented Alda by extending the DistAlgo
compiler~\cite{Liu+12DistPL-OOPSLA,Liu+17DistPL-TOPLAS,distalgo22github}.
%
DistAlgo is an extension of Python with high-level set queries as well as
distributed processes.  The 
compiler is implemented in Python~3, and uses the Python parser.  
So Python syntax is used in place of the ideal syntax presented in
Section~\ref{sec-lang}, allowing any user with Python to run Alda directly.
 

The Alda implementation extends the DistAlgo compiler to support rule-set
definitions, function \co{\INFER}, and maintenance of derived predicates at updates to non-local variables.
It handles direct updates to variables used as predicates, 
not updates through aliasing, as
we found this to be the only update case in all benchmarks and other examples
we have seen; we think this is because using logic rules with updates is 
similar to using queries and updates in relational databases, with no need
of updates through aliasing.
%
Currently Datalog rules extended with unrestricted negation are supported,
and well-founded semantics computed by XSB is used;
extensions for more general rules can be handled similarly,
and inference using XSB can remain the same.
Calls to \co{\INFER} are automatically added at updates to non-local base
predicates of rule sets.
%

In particular, the following Python syntax is used for rule sets, where a
rule can be either one of the two forms below, so the only restriction is
that the name \co{rules} 
is reserved.\noticlp{\Vex{-.255}}

\begin{code}
  def rules (name = \p{rsname}):
    \p{conclusion}, if_(\p{hypothesis\sb{1}}, \p{hypothesis\sb{2}}, \p{...}, \p{hypothesis\sb{h}})
    if (\p{hypothesis\sb{1}}, \p{hypothesis\sb{2}}, \p{...}, \p{hypothesis\sb{h}}): \p{conclusion}
    \p{...}
\end{code}\Vex{-.5}
Rule sets are translated into Prolog rules at compile time.  
The directive \co{:-~auto\_table.} is added for automatic tabling
in XSB.

For function \co{\INFER}, the implementation translates the values of
predicates and the list of queries into facts and queries in standard
Prolog syntax, and translates the query answers back to values of set
variables.  It invokes XSB using a command line in between, passing data
through files; this external interface has an obvious overhead, but it has
not affected Alda having generally good performance.  \co{\INFER}
automatically reads and writes non-local predicates used in a rule set.

Note that the overhead of the external interface can be removed with an
in-memory interface from Python to XSB, which is actively being developed
by the XSB team.\footnote{A version\noticlp{ working} for Unix, not yet Windows, has
  been released: passing data of size 100 million in memory took
  about 30 nanoseconds per element~\cite[release notes]{xsb22}.  So even
  the largest data in our experiments, of size a few millions, would
  take 0.1--0.2 seconds to pass in memory, instead of 10--20 seconds with the
  current external interface.}
However, even with the overhead of the external interface, Alda is still
faster or even drastically faster than half or more of the rule engines
tested in OpenRuleBench~\cite{Lia+09open} for all benchmarks measured
except DBLP (even though OpenRuleBench uses the fastest manually optimized
program for each problem for each rule engine), and than not using rules at
all (without manually writing or adapting a drastically more complex,
specialized algorithm implementation for each problem).

\notes{
With the \extern call interface implemented in XSB\footnote{Thanks to David
  Warren for the implementation as a 28-line XSB program.}, 
our compiler only needs to issue the following 
command line after writing the rules and facts files and before reading the
answers file:
\begin{code}
  xsb -e "external\_file\_query(filename, query)."
\end{code}
We generate \co{:- auto\_table.} at the top of the rules file,
so all the rules are compiled with automatic tabling in XSB~\cite{CheWar96}.
}




Building on top of DistAlgo and XSB, the compiler consists of about 
1100
lines of Python and about 50 lines of XSB.
This is owing critically to the overall framework and comprehensive 
support,
especially 
for high-level queries, already in the DistAlgo compiler and to the 
powerful query 
engine of XSB.  The parser for the rule extension is
about 270 
lines, and 
update analysis and code generation for rules and inference are about 800
lines.

The current compiler does not perform further optimizations, because they
are orthogonal to the focus of this paper, and our experiments already
showed generally good performance.  Further optimizations can be
implemented in either the Alda compiler to generate optimized rules
and tabling and indexing directives, or in XSB.  Incremental maintenance
under updates can also be implemented in either one, with a slightly richer
interface between the two.
%

\iclp{
\begin{table}[t]
\centering
\footnotesize
\begin{tabular}{p{13ex}||p{\noticlp{18.5}\iclp{19}ex}|p{\noticlp{12.5}\iclp{13}ex}||@{~}p{22ex}||@{~}p{17.5ex}}
Benchmark \mbox{}~~~~~sets & ~~~Benchmarks & ~~Variants\newline \mbox{}\,and timing & ~~\,Problem kinds & Code/data size\\\noticlp{\hline}\iclp{\cline{1-5}}
Open- RuleBench \cite{Lia+09open}
    & 13 incl.\ 
        LUBM,\newline Mondial, DBLP,\newline TC, 
        WordNet,\newline Wine
    & TCrev,\newline TCda,\newline TCpy,\newline ORBtimer 
    & many kinds of\newline rules and queries,\newline 
        but missing\newline aggregate queries
    & largest rule set:\newline 967 rules,\newline largest~data~size:\newline 2.4M+\\\noticlp{\hline}\iclp{\cline{1-5}}
RBAC\newline as in\newline Section~\ref{sec-obj}
    & RBACallloc,\newline RBACnonloc,\newline RBACunion
    & RBACda, RBACpy, RBACtimer 
    & interleaved object\newline queries and updates\newline 
        with function and \newline recursive rules
    & program size:\newline 385--423,\newline data size:\newline 10K+ \\\noticlp{\hline}\iclp{\cline{1-5}}
Program Analysis
    & PA~(on~any~prog.:\newline numpy, pandas,\newline matplot,~pytorch,\newline sympy, etc.) 
    & PAopt,\newline PAtimer 
    & interleaved rules,\newline aggregate and set\newline queries, and\newline recursive functions
    & program size:\newline 55\,XSB,\,33\,Alda,\newline largest~data~size:\newline 5.1M+ \\\noticlp{\hline}\iclp{\cline{1-5}}
\end{tabular}\Vex{1}
\caption{Benchmarks from different kinds of problems. RBAC benchmarks are for different ways of using rules as at the end of Section~\ref{sec-infer}. PA is a mixture of problems from class hierarchy analysis. 
Under Variants, suffixes py and da indicate using \co{while} loops like that in Section~\ref{sec-pred} in Python and DistAlgo, respectively, instead of using rules.}
\label{tab-benches}
\end{table}
}
We discuss our experiments on the benchmarks 
\noticlp{described in Section~\ref{sec-bench}}\iclp{summarized in Table~\ref{tab-benches}}.
\iclp{Detailed description of the benchmarks are in~\cite{Liu+22RuleLang-arxiv,Liu+23RuleLangBench-ICLP}.}
Just as the benchmarks selected, the experiments selected are also meant to
show generally good performance even under the most extreme overhead penalties
we have encountered---runs with large data \iclp{(DBLP and PA)}, large query results \iclp{(transitive closure TC)}, large
rules \iclp{(Wine)}, frequent switches among different ways of using rules and other
features \iclp{(RBAC and PA)}, and frequent external invocations of the rule engine \iclp{(RBAC)}.  Our
extensive experiments with other uses of Alda
have experienced minimum performance overhead.

All measurements were taken on
a machine with an Intel Xeon X5690 3.47 GHz CPU, 94 GB RAM,
running 64-bit Ubuntu 16.04.7,
Python 3.9.9, and XSB 4.0.0.
%
For each experiment, the reported running times are CPU times averaged over
10 runs.  Garbage collection in Python was disabled for smoother running
times when calling XSB.
Program sizes are numbers of lines excluding comments and empty lines.
Data sizes are number of facts.\notes{said already}

We summarize the results from the experiments below. Detailed measurements
and explanations are in~\cite{Liu+22RuleLang-arxiv,Liu+23RuleLangBench-ICLP}.

\noticlp{
\mypar{Compilation times and program size}
The Alda programs are 4--970 lines for benchmarks from OpenRuleBench,
385--423 lines for RBAC benchmarks, and 33 lines for program analysis
(PA) benchmarks.
For each set of benchmarks (OpenRuleBench, RBAC, and PA), a single small
shared Alda file (40--45 lines) for benchmarking is used.

The compilation times for all programs are 0.6 seconds or less, and for all
but RBAC benchmarks and Wine in OpenRulBench are about 0.1 seconds or less.
The generated Python files are 16--538 lines: the largest ones for RBAC
benchmarks, 93 lines for PA benchmarks, and all fewer than 50 lines for
benchmarks from OpenRuleBench.
The generated XSB files are all smaller than the Alda files, because they
contain only the rules, from just a few lines for many benchmarks to 968
lines for Wine from OpenRuleBench.

Note that for Alda programs that have corresponding XSB programs
(OpenRuleBench in Table~\ref{tab-rulebench} and PA), Alda programs are all
much smaller, almost all by dozens or even hundreds of lines, and by an
order of magnitude for Join1 and TC in OpenRuleBench, because we have all
the benchmarking code in the shared benchmarking file.

\mypar{Performance of classical queries using rules}
For the classical transitive closure problem, we experimented with TC from
OpenRuleBench, a well-known variant with the two predicates reversed in the
recursive rule, and the Python and Alda \co{\WHILE} loops in
Sections~\ref{sec-prob} and \ref{sec-pred}, respectively.  The same data
generation as OpenRuleBench is used to return large query results---almost
complete graphs.

The results are as expected: the two Alda programs that use rules are
asymptotically, drastically faster than Python and Alda \co{\WHILE} loops,
and they exhibit known notable performance
differences~\cite{TekLiu10RuleQuery-PPDP,TekLiu11RuleQueryBeat-SIGMOD}.
Most notably but as expected, passing the query results back from XSB has a
high overhead, up to 5.9 seconds, out of 29.2 seconds total, for graphs of 100K
edges, but this overhead is expected to be reduced to 1\% of it when the
in-memory Python-XSB interface is used.

\mypar{Integrating with objects, updates, and set queries}
We use RBAC benchmarks in Table~\ref{tab-rbac}, Section~\ref{sec-rbac}, for
this evaluation, especially with frequent queries and updates and
intensively frequent restart of XSB for queries randomly mixed with updates
of the queried data: 5000 users, 500 roles, 5000 UR relation, 550 RH
relation, up to 500 queries, and 230 updates of various kinds.

The results are as expected as well: RBAunion and RBACalloc are very close,
and are much slower than RBACnonloc---up to 331.7 and 333.9 seconds, respectively, vs.\
97.9 seconds.
Most notably but as expected, the overhead of repeated queries using XSB is
high for RBACunion and RBACalloc, but low for RBACnonloc, up to 134.5 and 145.9
seconds, respectively, vs.\ 2.8 seconds. The highest overhead is from restarting XSB 500 times,
which will be totally eliminated when in-memory interface is used.

\mypar{Integrating with aggregate queries and recursive functions}
We use PA and PAopt benchmarks and their corresponding programs in XSB, 
as described in Section~\ref{sec-pa}, for this evaluation, 
and we focus on applying the analysis to large programs as input data.  The
programs analyzed include 9 widely-used open-source Python packages for a
wide range of application domains: NumPy, SciPy, MatPlotLib, SymPy, Pandas,
SymPy, Django, Scikit-learn, and PyTorch---with 641K--5.1M input facts
total and 252K--2.2M facts used by the analysis.

The results are not as expected: we found the corresponding XSB
programs to be highly inefficient, being all slower and even drastically
slower than Alda programs, even 120 times slower for PyTorch.
Significant effort was spent on performance debugging and manual
optimization, and we eventually created a version that is faster than Alda---5.1
seconds vs.\ 15.2 seconds on the largest input, SymPy---by using additional
directives for targeted tabling that also subsumes some indexing.

As expected, the Alda programs here have a high overhead of 
passing the data to XSB, up to 13.1 seconds on SymPy, which again is expected to be reduced
to 1\% of it with in-memory Python-XSB interface. This means that the
resulting Alda programs would be faster than even the manually optimized XSB, showing that computations
not using rules, e.g., aggregations and functions, are not only simpler and
easier in Alda/Python than in XSB but also faster.

\mypar{Scaling with data and rules}
We use the two largest benchmarks from OpenRuleBench: DBLP, with over 2.4 million facts,
the largest real-world data set among all in OpenRuleBench; and Wine, with
961 rules, the largest rule set among all. The results are again as expected.  

For DBLP, XSB is more than three times as fast as Alda, 
9.5 seconds vs.\ 30.6 seconds, but as for PA benchmarks, the overhead of passing the
large data to XSB is large, here 26.9 seconds, and is expected to be reduced to 1\% of that; note that the Alda program has faster reading from
pickled data.

For Wine, XSB is more than eight times as fast as Alda, 3.8 seconds vs.\ 31.0
seconds.  This is due to the use of \co{auto\_table} in Alda generated code,
which does variant tabling, whereas this program, through manual debugging
and optimization, was found to need subsumptive
tabling~\cite{TekLiu11RuleQueryBeat-SIGMOD}.
Optimizations~\cite{LiuSto09Rules-TOPLAS,TekLiu10RuleQuery-PPDP,TekLiu11RuleQueryBeat-SIGMOD}
can be added to the Alda compiler to match this efficiency automatically.
Note that this slowest Alda program is still faster than half of
the systems tested in OpenRuleBench, which took up to 140 seconds and 
three systems gave errors, and where XSB was the fastest at 4.47 seconds~\cite{Lia+09open}.

}
\iclp{
\begin{itemize}
\item Compared with XSB programs in OpenRuleBench, the corresponding Alda programs are much smaller, almost all by dozens or even hundreds of lines, because all benchmarking code is in a single shared 45-line ORBtimer, much easier in Python than XSB.  Compilation times are all 0.6 seconds or less.

\item Running times for all benchmarks and variants, except for PA, are as expected, e.g., TC is drastically faster than TCpy and TCda, and essentially as fast as XSB if not for the overhead of using external interface with XSB; and RBACnonloc is much faster than RBACallloc due to updates being much less frequent than queries. 
The overhead of using external interface is obvious:
e.g., for TC, up to 5.9 seconds, out of 29.2, for graphs of 100K edges; 
for PA, 13.1 seconds, out of 15.2, on the largest program, SymPy;
and worst for DBLP, 26.9 seconds, out of 30.6, on over 2.4M facts. 

However, even so, Alda is competitive, as described above, and the overhead is expected to be reduced to 1\% of it with an in-memory Python-XSB interface.

\item For PA, the corresponding XSB programs were all slower and even drastically
slower than Alda programs, even 120 times slower on PyTorch.
Significant effort was spent on performance debugging and manual
optimization before we eventually created a version that is faster than Alda---5.1
vs.\ 15.2 seconds on SymPy.
\end{itemize}
}

\mysec{Related work and conclusion}
\label{sec-related}

%
%
%
%
%
%

There has been extensive effort in design and implementation of
languages to support programming with logic rules together with other
programming paradigms, by extending logic languages, extending languages in
other paradigms, or developing multi-paradigm or other standalone
languages.

A large variety of logic rule languages have been extended to support sets,
functions, updates, and/or objects,
etc.~\cite{KifLiu18wbook,prolog50tplp}.  For example, see Maier
et al.~\cite{maier18hist-wbook} for Datalog and variants extended with
sets, functions, objects, updates, higher-order extensions, and more.
In particular, many Prolog variants support sets, functions, updates, 
objects, constraints, etc.  For example, Prolog supports \co{assert} for
updates, as well as cut and negation as failure that are imperative instead
of declarative~\cite{Sterling:Shapiro:94};
Flora~\cite{YanKif00flora,flora20} builds on XSB and supports objects
(F-logic), higher-order programming (HiLog), and updates (Transaction
Logic); and Picat~\cite{Zhou16picat} builds on B-Prolog and
supports updates, comprehensions, etc.
Lambda Prolog~\cite{miller2012programming} extends Prolog with 
simply typed lambda terms and higher-order programming.
Functional logic languages, such as Mercury~\cite{somogyi1995mercury} 
and Curry~\cite{hanus2013functional}, 
combine functional programming and logic programming.
Some logic programming systems are driven by scripting externally, e.g.,
using Lua for IDP~\cite{bru14predicate}, and shell scripts for
LogicBlox~\cite{aref2015design}.
Additional examples of Datalog extensions include Flix~\cite{madsen2016datalog,madsen2020fixpoints}, which 
supports lattices and monotone functions, 
and DDlog~\cite{ryzhyk2019differential}, which 
supports incremental maintenance under updates to input relations. 
%
These languages and extensions do not support predicates as set-valued 
variables together with commonly-used updates and objects
in a simple and direct way, 
or do not support them at all.  

Many languages in other programming paradigms, especially including imperative
languages and object-oriented languages, have been extended to support
rules by being a host language.
This is generally through explicit library interfaces of the host languages
to connect with a particular logic
language, for example, a Java interface for XSB through
InterProlog~\cite{calejo04inteprolog,xsb22}, C++ and Python interfaces for
answer-set programming systems dlvhex~\cite{Redl16} and
Potassco~\cite{ban17clingcon}, a Python interface for
IDP~\cite{ven17idp-py}, Rust and other interfaces for DDlog~\cite{ryzhyk2019differential}, and many more, e.g., for miniKanren~\cite{byrd2009relational}.
%
Hosting logic languages through 
explicit interfaces requires programmers to write 
extra wrapper code for going to the rule language and coming back---declare predicates and/or logic variables, wrap 
features in special objects, functions, macros, etc., 
and/or convert data to and from special representations.
They are in the same spirit as interfaces such as JDBC~\cite{reese2000database} for
using database systems from languages such as Java.

Multi-paradigm languages and other standalone languages 
have also been developed. 
For example, the Mozart 
system for the Oz multi-paradigm programming
language~\cite{RoyHar04} supports logic, functional, and
constraint as well as imperative and concurrent programming. However, it is
similar to logic languages extended with other features, because it supports
logic variables, but not state variables to be assigned to as in commonly-used imperative
languages.
Examples of other languages involving logic and constraints with updates
and/or objects include LOGRES~\cite{cacace1990integrating}, which
integrates object-oriented data modeling and updates with rules under
inflationary semantics; TLA+~\cite{lam94tla}, a logic language for
specifying actions;
CLAIRE~\cite{caseau2002claire}, an object-oriented language that supports
functions, sets, and rules whose conclusions are actions;
LINQ~\cite{meijer2006linq,LINQ}, an extension of C\# for SQL-like queries;
IceDust~\cite{harkes16IceDust}, a Java-based language for querying data
with path-based navigation and incremental computation;
extended LogiQL in SolverBlox~\cite{aref18solverblox-wbook}, for
mathematical and logic programming on top of Datalog with updates and
constraints; and other logic-based query languages,
e.g.,~Datomic~\cite{anderson2016datomic} and SOUL~\cite{de2011soul}.
These are either logic languages lacking general imperative and
objected-oriented programming constructs, or imperative and object-oriented
languages lacking the power and full declarativeness of logic rules.

%
In conclusion, Alda supports ease of programming with
logic rules together with all of sets,
functions, updates, and objects as seamlessly integrated built-ins,
without extra interfaces or boiler-plate code.
As a direction for future work, many optimizations can be added to improve
the efficiency of implementations.
This includes optimizing the logic rule engines used~\cite{LiuSto09Rules-TOPLAS,TekLiu11RuleQueryBeat-SIGMOD}, the interfaces
and interactions with them, and using other efficient rule systems such as
Clingo~\cite{gebser2019multi} and specialized rule implementations
such as Souffle~\cite{jordan2016souffle} 
to obtain the best possible performance.
%


\noticlp{\mypar{\large Acknowledgments}}\iclp{\Vex{-2}\subsection*{Acknowledgments}}\Vex{-.5}
We thank David S.\ Warren for an initial 28-line XSB program for interface
to XSB, and Tuncay Tekle for help implementing some benchmarks and running some preliminary experiments.
We also thank 
Thang Bui for additional applications in program analysis and optimization, 
and students in undergraduate and graduate courses
for using Alda and its earlier versions, called DA-rules.

\iclp{\Vex{-2}}
{
\renewcommand{\baselinestretch}{-.1}\small
\def\usebib{
\def\bibdir{../../../bib}   
{
\bibliography{\bibdir/strings,\bibdir/liu,\bibdir/IC,\bibdir/PT,\bibdir/PA,\bibdir/Lang,\bibdir/Algo,\bibdir/DB,\bibdir/AI,\bibdir/Sec,\bibdir/Sys,\bibdir/SE,\bibdir/Vis,\bibdir/misc,\bibdir/crossref} 

\begin{thebibliography}{}

\bibitem[\protect\citeauthoryear{Abiteboul, Hull, and Vianu}{Abiteboul
  et~al\mbox{.}}{1995}]{AbiHulVia95}
{\sc Abiteboul, S.}, {\sc Hull, R.}, {\sc and} {\sc Vianu, V.} 1995.
\newblock {\em Foundations of Databases: The Logical Level}.
\newblock Addison-Wesley.

\bibitem[\protect\citeauthoryear{Anderson, Gaare, Holgu\'in, Bailey, and
  Pratley}{Anderson et~al\mbox{.}}{2016}]{anderson2016datomic}
{\sc Anderson, J.}, {\sc Gaare, M.}, {\sc Holgu\'in, J.}, {\sc Bailey, N.},
  {\sc and} {\sc Pratley, T.} 2016.
\newblock The {Datomic} database.
\newblock In {\em Professional Clojure}. Wiley Online Library, Chapter~6,
  169--215.

\bibitem[\protect\citeauthoryear{{ANSI INCITS}}{{ANSI
  INCITS}}{2004}]{ansi04role}
{\sc {ANSI INCITS}}. 2004.
\newblock {Role-Based Access Control}.
\newblock ANSI INCITS 359-2004, American National Standards Institute,
  International Committee for Information Technology Standards.

\bibitem[\protect\citeauthoryear{Aref, ten Cate, Green, Kimelfeld, Olteanu,
  Pasalic, Veldhuizen, and Washburn}{Aref et~al\mbox{.}}{2015}]{aref2015design}
{\sc Aref, M.}, {\sc ten Cate, B.}, {\sc Green, T.~J.}, {\sc Kimelfeld, B.},
  {\sc Olteanu, D.}, {\sc Pasalic, E.}, {\sc Veldhuizen, T.~L.}, {\sc and} {\sc
  Washburn, G.} 2015.
\newblock Design and implementation of the {LogicBlox} system.
\newblock In {\em Proceedings of the 2015 ACM SIGMOD International Conference
  on Management of Data}. 1371--1382.
\newblock \url{https://doi.org/10.1145/2723372.2742796}.

\bibitem[\protect\citeauthoryear{Banbara, Kaufmann, Ostrowski, and
  Schaub}{Banbara et~al\mbox{.}}{2017}]{ban17clingcon}
{\sc Banbara, M.}, {\sc Kaufmann, B.}, {\sc Ostrowski, M.}, {\sc and} {\sc
  Schaub, T.} 2017.
\newblock Clingcon: {The} next generation.
\newblock {\em Theory and Practice of Logic Programming\/}~{\em 17,\/}~4,
  408--461.

\bibitem[\protect\citeauthoryear{Bancilhon, Maier, Sagiv, and Ullman}{Bancilhon
  et~al\mbox{.}}{1986}]{Ban+86}
{\sc Bancilhon, F.}, {\sc Maier, D.}, {\sc Sagiv, Y.}, {\sc and} {\sc Ullman,
  J.~D.} 1986.
\newblock Magic sets and other strange ways to implement logic programs.
\newblock In {\em Proceedings of the 5th ACM SIGACT-SIGMOD Symposium on
  Principles of Database Systems}. 1--16.

\bibitem[\protect\citeauthoryear{Borraz-S\'anchez, Klabjan, Pasalic, and
  Aref}{Borraz-S\'anchez et~al\mbox{.}}{2018}]{aref18solverblox-wbook}
{\sc Borraz-S\'anchez, C.}, {\sc Klabjan, D.}, {\sc Pasalic, E.}, {\sc and}
  {\sc Aref, M.} 2018.
\newblock {SolverBlox: Algebraic} modeling in {Datalog}.
\newblock In {\em Declarative Logic Programming: Theory, Systems, and
  Applications}, {M.~Kifer} {and} {Y.~A. Liu}, Eds. ACM and Morgan \& Claypool,
  Chapter~6, 331--356.
\newblock \url{https://doi.org/10.1145/3191315.3191322}.

\bibitem[\protect\citeauthoryear{Bruynooghe, Blockeel, Bogaerts, De~Cat,
  De~Pooter, Jansen, Labarre, Ramon, Denecker, and Verwer}{Bruynooghe
  et~al\mbox{.}}{2014}]{bru14predicate}
{\sc Bruynooghe, M.}, {\sc Blockeel, H.}, {\sc Bogaerts, B.}, {\sc De~Cat, B.},
  {\sc De~Pooter, S.}, {\sc Jansen, J.}, {\sc Labarre, A.}, {\sc Ramon, J.},
  {\sc Denecker, M.}, {\sc and} {\sc Verwer, S.} 2014.
\newblock Predicate logic as a modeling language: {M}odeling and solving some
  machine learning and data mining problems with {IDP3}.
\newblock {\em Theory and Practice of Logic Programming\/}~{\em 15,\/}~6,
  783--817.
\newblock \url{https://doi.org/10.1017/S147106841400009X}.

\bibitem[\protect\citeauthoryear{Byrd}{Byrd}{2009}]{byrd2009relational}
{\sc Byrd, W.~E.} 2009.
\newblock Relational programming in {miniKanren}: Techniques, applications, and
  implementations.
\newblock Ph.D. thesis, Indiana University.

\bibitem[\protect\citeauthoryear{Cacace, Ceri, Crespi-Reghizzi, Tanca, and
  Zicari}{Cacace et~al\mbox{.}}{1990}]{cacace1990integrating}
{\sc Cacace, F.}, {\sc Ceri, S.}, {\sc Crespi-Reghizzi, S.}, {\sc Tanca, L.},
  {\sc and} {\sc Zicari, R.} 1990.
\newblock Integrating object-oriented data modelling with a rule-based
  programming paradigm.
\newblock In {\em Proceedings of the 1990 ACM SIGMOD international conference
  on Management of data}. 225--236.

\bibitem[\protect\citeauthoryear{Calejo}{Calejo}{2004}]{calejo04inteprolog}
{\sc Calejo, M.} 2004.
\newblock {InterProlog}: {Towards} a declarative embedding of logic programming
  in {Java}.
\newblock In {\em Proceedings of the 9th European Conference on Logics in
  Artificial Intelligence}. LNCS, vol. 3229. Springer, 714--717.

\bibitem[\protect\citeauthoryear{Caseau, Josset, and Laburthe}{Caseau
  et~al\mbox{.}}{2002}]{caseau2002claire}
{\sc Caseau, Y.}, {\sc Josset, F.-X.}, {\sc and} {\sc Laburthe, F.} 2002.
\newblock Claire: Combining sets, search and rules to better express
  algorithms.
\newblock {\em Theory and Practice of Logic Programming\/}~{\em 2,\/}~6,
  769--805.

\bibitem[\protect\citeauthoryear{Chen and Warren}{Chen and
  Warren}{1996}]{CheWar96}
{\sc Chen, W.} {\sc and} {\sc Warren, D.~S.} 1996.
\newblock Tabled evaluation with delaying for general logic programs.
\newblock {\em Journal of the ACM\/}~{\em 43,\/}~1, 20--74.

\bibitem[\protect\citeauthoryear{De~Roover, Noguera, Kellens, and
  Jonckers}{De~Roover et~al\mbox{.}}{2011}]{de2011soul}
{\sc De~Roover, C.}, {\sc Noguera, C.}, {\sc Kellens, A.}, {\sc and} {\sc
  Jonckers, V.} 2011.
\newblock The {SOUL} tool suite for querying programs in symbiosis with
  {Eclipse}.
\newblock In {\em Proceedings of the 9th International Conference on Principles
  and Practice of Programming in Java}. 71--80.

\bibitem[\protect\citeauthoryear{Ferraiolo, Sandhu, Gavrila, Kuhn, and
  Chandramouli}{Ferraiolo et~al\mbox{.}}{2001}]{ferraiolo01proposed}
{\sc Ferraiolo, D.~F.}, {\sc Sandhu, R.}, {\sc Gavrila, S.}, {\sc Kuhn, D.~R.},
  {\sc and} {\sc Chandramouli, R.} 2001.
\newblock Proposed {NIST} standard for role-based access control.
\newblock {\em ACM Transactions on Information and Systems Security\/}~{\em
  4,\/}~3, 224--274.

\bibitem[\protect\citeauthoryear{Fitting}{Fitting}{2002}]{fitting2002fixpoint}
{\sc Fitting, M.} 2002.
\newblock Fixpoint semantics for logic programming: {A} survey.
\newblock {\em Theoretical Computer Science\/}~{\em 278,\/}~1, 25--51.

\bibitem[\protect\citeauthoryear{Fong and Ullman}{Fong and
  Ullman}{1976}]{FonUll76}
{\sc Fong, A.~C.} {\sc and} {\sc Ullman, J.~D.} 1976.
\newblock Inductive variables in very high level languages.
\newblock In {\em Conference Record of the 3rd Annual ACM Symposium on
  Principles of Programming Languages}. 104--112.

\bibitem[\protect\citeauthoryear{Gebser, Kaminski, Kaufmann, and Schaub}{Gebser
  et~al\mbox{.}}{2019}]{gebser2019multi}
{\sc Gebser, M.}, {\sc Kaminski, R.}, {\sc Kaufmann, B.}, {\sc and} {\sc
  Schaub, T.} 2019.
\newblock Multi-shot {ASP} solving with clingo.
\newblock {\em Theory and Practice of Logic Programming\/}~{\em 19,\/}~1,
  27--82.
\newblock \url{https://doi.org/10.1017/S1471068418000054}.

\bibitem[\protect\citeauthoryear{Geiger}{Geiger}{1995}]{geiger1995inside}
{\sc Geiger, K.} 1995.
\newblock {\em Inside ODBC}.
\newblock Microsoft Press.

\bibitem[\protect\citeauthoryear{Gorbovitski, Liu, Stoller, Rothamel, and
  Tekle}{Gorbovitski et~al\mbox{.}}{2010}]{Gor+10Alias-DLS}
{\sc Gorbovitski, M.}, {\sc Liu, Y.~A.}, {\sc Stoller, S.~D.}, {\sc Rothamel,
  T.}, {\sc and} {\sc Tekle, T.} 2010.
\newblock Alias analysis for optimization of dynamic languages.
\newblock In {\em Proceedings of the 6th Symposium on Dynamic Languages}. ACM
  Press, 27--42.
\newblock \url{https://doi.org/10.1145/1869631.1869635}.

\bibitem[\protect\citeauthoryear{Goyal}{Goyal}{2005}]{Goy05}
{\sc Goyal, D.} 2005.
\newblock Transformational derivation of an improved alias analysis algorithm.
\newblock {\em Higher-Order and Symbolic Computation\/}~{\em 18,\/}~1--2,
  15--49.

\bibitem[\protect\citeauthoryear{Gupta and Mumick}{Gupta and
  Mumick}{1999}]{GupMum99maint}
{\sc Gupta, A.} {\sc and} {\sc Mumick, I.~S.} 1999.
\newblock Maintenance of materialized views: {P}roblems, techniques, and
  applications.
\newblock In {\em Materialized Views: Techniques, Implementations, and
  Applications}. MIT Press, 145--157.

\bibitem[\protect\citeauthoryear{Hanus}{Hanus}{2013}]{hanus2013functional}
{\sc Hanus, M.} 2013.
\newblock Functional logic programming: From theory to {Curry}.
\newblock In {\em Programming Logics}. Springer, 123--168.

\bibitem[\protect\citeauthoryear{Harkes, Groenewegen, and Visser}{Harkes
  et~al\mbox{.}}{2016}]{harkes16IceDust}
{\sc Harkes, D.~C.}, {\sc Groenewegen, D.~M.}, {\sc and} {\sc Visser, E.} 2016.
\newblock {IceDust: Incremental and eventual computation of derived values in
  persistent object graphs}.
\newblock In {\em 30th European Conference on Object-Oriented Programming}.
  LIPIcs, vol.~56. Schloss Dagstuhl--Leibniz-Zentrum fuer Informatik,
  11:1--11:26.

\bibitem[\protect\citeauthoryear{Jaeger and Tidswell}{Jaeger and
  Tidswell}{2000}]{Jaeger:Tidswell:00}
{\sc Jaeger, T.} {\sc and} {\sc Tidswell, J.} 2000.
\newblock Rebuttal to the {NIST} {RBAC} model proposal.
\newblock In {\em Proceedings of the 5th ACM Workshop on Role Based Access
  Control}. 66.

\bibitem[\protect\citeauthoryear{Jordan, Scholz, and Suboti{\'c}}{Jordan
  et~al\mbox{.}}{2016}]{jordan2016souffle}
{\sc Jordan, H.}, {\sc Scholz, B.}, {\sc and} {\sc Suboti{\'c}, P.} 2016.
\newblock Souffl{\'e}: {On} synthesis of program analyzers.
\newblock In {\em Proceedings of the International Conference on Computer Aided
  Verification}. Springer, 422--430.

\bibitem[\protect\citeauthoryear{Kifer and Liu}{Kifer and
  Liu}{2018}]{KifLiu18wbook}
{\sc Kifer, M.} {\sc and} {\sc Liu, Y.~A.}, Eds. 2018.
\newblock {\em Declarative Logic Programming: Theory, Systems, and
  Applications}.
\newblock ACM and Morgan \& Claypool.

\bibitem[\protect\citeauthoryear{Kifer, Yang, Wan, and Zhao}{Kifer
  et~al\mbox{.}}{2020}]{flora20}
{\sc Kifer, M.}, {\sc Yang, G.}, {\sc Wan, H.}, {\sc and} {\sc Zhao, C.} 2020.
\newblock {\em Ergo Lite (a.k.a.\ Flora-2): User's Manual Version 2.1}.
\newblock Stony Brook University.
\newblock \url{http://flora.sourceforge.net/}. Accessed May 25, 2023.

\bibitem[\protect\citeauthoryear{K\"orner, Leuschel, Barbosa, Costa, Dahl,
  Hermenegildo, Morales, Wielemaker, Diaz, Abreu, and Ciatto}{K\"orner
  et~al\mbox{.}}{2022}]{prolog50tplp}
{\sc K\"orner, P.}, {\sc Leuschel, M.}, {\sc Barbosa, J.~a.}, {\sc Costa,
  V.~S.}, {\sc Dahl, V.}, {\sc Hermenegildo, M.~V.}, {\sc Morales, J.~F.}, {\sc
  Wielemaker, J.}, {\sc Diaz, D.}, {\sc Abreu, S.}, {\sc and} {\sc Ciatto, G.}
  2022.
\newblock Fifty years of {Prolog} and beyond.
\newblock {\em Theory and Practice of Logic Programming\/}~{\em 22,\/}~6,
  776--858.
\newblock \url{https://doi.org/10.1017/S1471068422000102}.

\bibitem[\protect\citeauthoryear{Lamport}{Lamport}{1994}]{lam94tla}
{\sc Lamport, L.} 1994.
\newblock The temporal logic of actions.
\newblock {\em ACM Transactions on Programming Languages and Systems\/}~{\em
  16,\/}~3, 872--923.

\bibitem[\protect\citeauthoryear{Li, Byun, and Bertino}{Li
  et~al\mbox{.}}{2007}]{Li+07critique}
{\sc Li, N.}, {\sc Byun, J.-W.}, {\sc and} {\sc Bertino, E.} 2007.
\newblock A critique of the {ANSI} standard on role-based access control.
\newblock {\em IEEE Security and Privacy\/}~{\em 5,\/}~6, 41--49.

\bibitem[\protect\citeauthoryear{Liang, Fodor, Wan, and Kifer}{Liang
  et~al\mbox{.}}{2009}]{Lia+09open}
{\sc Liang, S.}, {\sc Fodor, P.}, {\sc Wan, H.}, {\sc and} {\sc Kifer, M.}
  2009.
\newblock {OpenRuleBench}: {A}n analysis of the performance of rule engines.
\newblock In {\em Proceedings of the 18th International Conference on World
  Wide Web}. ACM Press, 601--610.

\bibitem[\protect\citeauthoryear{Lin and Liu}{Lin and
  Liu}{2022}]{distalgo22github}
{\sc Lin, B.} {\sc and} {\sc Liu, Y.~A.} 2014 (Latest update January 30, 2022).
\newblock {DistAlgo}: A language for distributed algorithms.
\newblock \url{http://github.com/DistAlgo}.
\newblock Accessed May 25, 2023.

\bibitem[\protect\citeauthoryear{Microsoft}{{LINQ}}{2023}]{LINQ}
{LINQ} 2023.
\newblock {Language Integrated Query (LINQ)}.
\newblock \url{https://docs.microsoft.com/dotnet/csharp/linq}.
\newblock Accessed May 25, 2023.

\bibitem[\protect\citeauthoryear{Liu}{Liu}{2018}]{liu18LPappl-wbook}
{\sc Liu, Y.~A.} 2018.
\newblock Logic programming applications: What are the abstractions and
  implementations?
\newblock In {\em Declarative Logic Programming: Theory, Systems, and
  Applications}, {M.~Kifer} {and} {Y.~A. Liu}, Eds. ACM and Morgan \& Claypool,
  Chapter~10, 519--557.
\newblock Also \url{https://arxiv.org/abs/1802.07284}.

\bibitem[\protect\citeauthoryear{Liu, Brandvein, Stoller, and Lin}{Liu
  et~al\mbox{.}}{2016}]{Liu+16IncOQ-PPDP}
{\sc Liu, Y.~A.}, {\sc Brandvein, J.}, {\sc Stoller, S.~D.}, {\sc and} {\sc
  Lin, B.} 2016.
\newblock Demand-driven incremental object queries.
\newblock In {\em Proceedings of the 18th International Symposium on Principles
  and Practice of Declarative Programming}. ACM Press, 228--241.
\newblock \url{https://doi.org/10.1145/2967973.2968610}.

\bibitem[\protect\citeauthoryear{Liu and Stoller}{Liu and
  Stoller}{2007}]{LiuSto07RBAC-ONR}
{\sc Liu, Y.~A.} {\sc and} {\sc Stoller, S.~D.} 2007.
\newblock Role-based access control: {A} corrected and simplified
  specification.
\newblock In {\em Department of Defense Sponsored Information Security
  Research: {N}ew Methods for Protecting Against Cyber Threats}. Wiley,
  425--439.

\bibitem[\protect\citeauthoryear{Liu and Stoller}{Liu and
  Stoller}{2009}]{LiuSto09Rules-TOPLAS}
{\sc Liu, Y.~A.} {\sc and} {\sc Stoller, S.~D.} 2009.
\newblock From {Datalog} rules to efficient programs with time and space
  guarantees.
\newblock {\em ACM Transactions on Programming Languages and Systems\/}~{\em
  31,\/}~6, 1--38.
\newblock \url{https://doi.org/10.1145/1552309.1552311}.

\bibitem[\protect\citeauthoryear{Liu and Stoller}{Liu and
  Stoller}{2020}]{LiuSto20Founded-JLC}
{\sc Liu, Y.~A.} {\sc and} {\sc Stoller, S.~D.} 2020.
\newblock Founded semantics and constraint semantics of logic rules.
\newblock {\em Journal of Logic and Computation\/}~{\em 30,\/}~8 (Dec.),
  1609--1638.
\newblock Also \url{http://arxiv.org/abs/1606.06269}.

\bibitem[\protect\citeauthoryear{Liu and Stoller}{Liu and
  Stoller}{2021}]{LiuSto21LogicalConstraints-JLC}
{\sc Liu, Y.~A.} {\sc and} {\sc Stoller, S.~D.} 2021.
\newblock Knowledge of uncertain worlds: Programming with logical constraints.
\newblock {\em Journal of Logic and Computation\/}~{\em 31,\/}~1 (Jan.),
  193--212.
\newblock Also \url{https://arxiv.org/abs/1910.10346}.

\bibitem[\protect\citeauthoryear{Liu and Stoller}{Liu and
  Stoller}{2022}]{LiuSto22RuleAgg-JLC}
{\sc Liu, Y.~A.} {\sc and} {\sc Stoller, S.~D.} 2022.
\newblock Recursive rules with aggregation: A simple unified semantics.
\newblock {\em Journal of Logic and Computation\/}~{\em 32,\/}~8 (Dec.),
  1659--1693.
\newblock Also \url{http://arxiv.org/abs/2007.13053}.

\bibitem[\protect\citeauthoryear{Liu, Stoller, and Lin}{Liu
  et~al\mbox{.}}{2017}]{Liu+17DistPL-TOPLAS}
{\sc Liu, Y.~A.}, {\sc Stoller, S.~D.}, {\sc and} {\sc Lin, B.} 2017.
\newblock From clarity to efficiency for distributed algorithms.
\newblock {\em ACM Transactions on Programming Languages and Systems\/}~{\em
  39,\/}~3 (May), 12:1--12:41.
\newblock Also \url{http://arxiv.org/abs/1412.8461}.

\bibitem[\protect\citeauthoryear{Liu, Stoller, Lin, and Gorbovitski}{Liu
  et~al\mbox{.}}{2012}]{Liu+12DistPL-OOPSLA}
{\sc Liu, Y.~A.}, {\sc Stoller, S.~D.}, {\sc Lin, B.}, {\sc and} {\sc
  Gorbovitski, M.} 2012.
\newblock From clarity to efficiency for distributed algorithms.
\newblock In {\em Proceedings of the 27th ACM SIGPLAN Conference on
  Object-Oriented Programming, Systems, Languages and Applications}. 395--410.
\newblock \url{https://doi.org/10.1145/2384616.2384645}.

\bibitem[\protect\citeauthoryear{Liu, Stoller, Tong, Lin, and Tekle}{Liu
  et~al\mbox{.}}{2022}]{Liu+22RuleLang-arxiv}
{\sc Liu, Y.~A.}, {\sc Stoller, S.~D.}, {\sc Tong, Y.}, {\sc Lin, B.}, {\sc
  and} {\sc Tekle, K.~T.} 2022.
\newblock Programming with rules and everything else, seamlessly.
\newblock {\em Computing Research Repository\/}~{\em arXiv:2205.15204 [cs.PL]}.
\newblock \url{http://arxiv.org/abs/2205.15204}.

\bibitem[\protect\citeauthoryear{Liu, Stoller, Tong, and Tekle}{Liu
  et~al\mbox{.}}{2023}]{Liu+23RuleLangBench-ICLP}
{\sc Liu, Y.~A.}, {\sc Stoller, S.~D.}, {\sc Tong, Y.}, {\sc and} {\sc Tekle,
  K.~T.} 2023.
\newblock Benchmarking for integrating logic rules with everything else.
\newblock In {\em Proceedings of the 39th International Conference on Logic
  Programming (Technical Communications)}. Open Publishing Association.

\bibitem[\protect\citeauthoryear{Madsen and Lhot{\'a}k}{Madsen and
  Lhot{\'a}k}{2020}]{madsen2020fixpoints}
{\sc Madsen, M.} {\sc and} {\sc Lhot{\'a}k, O.} 2020.
\newblock Fixpoints for the masses: {Programming} with first-class {Datalog}
  constraints.
\newblock {\em Proceedings of the ACM on Programming Languages\/}~{\em
  4,\/}~OOPSLA, 1--28.

\bibitem[\protect\citeauthoryear{Madsen, Yee, and Lhot{\'a}k}{Madsen
  et~al\mbox{.}}{2016}]{madsen2016datalog}
{\sc Madsen, M.}, {\sc Yee, M.-H.}, {\sc and} {\sc Lhot{\'a}k, O.} 2016.
\newblock From {Datalog} to {Flix}: {A} declarative language for fixed points
  on lattices.
\newblock {\em ACM SIGPLAN Notices\/}~{\em 51,\/}~6, 194--208.

\bibitem[\protect\citeauthoryear{Maier, Tekle, Kifer, and Warren}{Maier
  et~al\mbox{.}}{2018}]{maier18hist-wbook}
{\sc Maier, D.}, {\sc Tekle, K.~T.}, {\sc Kifer, M.}, {\sc and} {\sc Warren,
  D.~S.} 2018.
\newblock Datalog: {Concepts}, history and outlook.
\newblock In {\em Declarative Logic Programming: Theory, Systems, and
  Applications}, {M.~Kifer} {and} {Y.~A. Liu}, Eds. ACM and Morgan \& Claypool,
  Chapter~1, 3--120.

\bibitem[\protect\citeauthoryear{Meijer, Beckman, and Bierman}{Meijer
  et~al\mbox{.}}{2006}]{meijer2006linq}
{\sc Meijer, E.}, {\sc Beckman, B.}, {\sc and} {\sc Bierman, G.} 2006.
\newblock {LINQ}: reconciling object, relations and {XML} in the {.NET}
  framework.
\newblock In {\em Proceedings of the 2006 ACM SIGMOD international conference
  on Management of data}. 706--706.

\bibitem[\protect\citeauthoryear{Miller and Nadathur}{Miller and
  Nadathur}{2012}]{miller2012programming}
{\sc Miller, D.} {\sc and} {\sc Nadathur, G.} 2012.
\newblock {\em Programming with Higher-Order Logic}.
\newblock Cambridge University Press.

\bibitem[\protect\citeauthoryear{Redl}{Redl}{2016}]{Redl16}
{\sc Redl, C.} 2016.
\newblock The {DLVHEX} system for knowledge representation: {{Recent}} advances
  (system description).
\newblock {\em Theory and Practice of Logic Programming\/}~{\em 16,\/}~5-6,
  866--883.

\bibitem[\protect\citeauthoryear{Reese}{Reese}{2000}]{reese2000database}
{\sc Reese, G.} 2000.
\newblock {\em Database Programming with JDBC and JAVA}.
\newblock O'Reilly Media, Inc.

\bibitem[\protect\citeauthoryear{Rothamel and Liu}{Rothamel and
  Liu}{2007}]{RotLiu07Retrieval-PEPM}
{\sc Rothamel, T.} {\sc and} {\sc Liu, Y.~A.} 2007.
\newblock Efficient implementation of tuple pattern based retrieval.
\newblock In {\em Proceedings of the ACM SIGPLAN 2007 Workshop on Partial
  Evaluation and Program Manipulation}. 81--90.
\newblock \url{https://doi.org/10.1145/1244381.1244394}.

\bibitem[\protect\citeauthoryear{Rothamel and Liu}{Rothamel and
  Liu}{2008}]{RotLiu08OSQ-GPCE}
{\sc Rothamel, T.} {\sc and} {\sc Liu, Y.~A.} 2008.
\newblock Generating incremental implementations of object-set queries.
\newblock In {\em Proceedings of the 7th International Conference on Generative
  Programming and Component Engineering}. ACM Press, 55--66.
\newblock \url{https://doi.org/10.1145/1449913.1449923}.

\bibitem[\protect\citeauthoryear{Roy and Haridi}{Roy and
  Haridi}{2004}]{RoyHar04}
{\sc Roy, P.~V.} {\sc and} {\sc Haridi, S.} 2004.
\newblock {\em Concepts, Techniques, and Models of Computer Programming}.
\newblock MIT Press.

\bibitem[\protect\citeauthoryear{Ryzhyk and Budiu}{Ryzhyk and
  Budiu}{2019}]{ryzhyk2019differential}
{\sc Ryzhyk, L.} {\sc and} {\sc Budiu, M.} 2019.
\newblock Differential datalog.
\newblock In {\em Datalog 2.0, 3rd International Workshop on the Resurgence of
  Datalog in Academia and Industry}. 56--67.

\bibitem[\protect\citeauthoryear{Sagonas, Swift, and Warren}{Sagonas
  et~al\mbox{.}}{1994}]{SagSW94xsb}
{\sc Sagonas, K.}, {\sc Swift, T.}, {\sc and} {\sc Warren, D.~S.} 1994.
\newblock {XSB} as an efficient deductive database engine.
\newblock In {\em Proceedings of the 1994 ACM SIGMOD International Conference
  on Management of Data}. ACM Press, 442--453.

\bibitem[\protect\citeauthoryear{Saha and Ramakrishnan}{Saha and
  Ramakrishnan}{2003}]{SahaRam03}
{\sc Saha, D.} {\sc and} {\sc Ramakrishnan, C.~R.} 2003.
\newblock Incremental evaluation of tabled logic programs.
\newblock In {\em Proceedings of the 19th International Conference on Logic
  Programming}. Springer, 392--406.
\newblock \url{https://doi.org/10.1007/978-3-540-24599-5_27}.

\bibitem[\protect\citeauthoryear{Sandhu, Ferraiolo, and Kuhn}{Sandhu
  et~al\mbox{.}}{2000}]{Sandhu+00}
{\sc Sandhu, R.}, {\sc Ferraiolo, D.}, {\sc and} {\sc Kuhn, R.} 2000.
\newblock The {NIST} model for role-based access control: {T}owards a unified
  standard.
\newblock In {\em Proceedings of the 5th ACM Workshop on Role-Based Access
  Control}. 47--63.

\bibitem[\protect\citeauthoryear{Serbanuta, Rosu, and Meseguer}{Serbanuta
  et~al\mbox{.}}{2009}]{serbanuta07rewriting}
{\sc Serbanuta, T.~F.}, {\sc Rosu, G.}, {\sc and} {\sc Meseguer, J.} 2009.
\newblock A rewriting logic approach to operational semantics.
\newblock {\em Information and Computation\/}~{\em 207}, 305--340.

\bibitem[\protect\citeauthoryear{Somogyi, Henderson, and Conway}{Somogyi
  et~al\mbox{.}}{1995}]{somogyi1995mercury}
{\sc Somogyi, Z.}, {\sc Henderson, F.~J.}, {\sc and} {\sc Conway, T.~C.} 1995.
\newblock Mercury, an efficient purely declarative logic programming language.
\newblock {\em Australian Computer Science Communications\/}~{\em 17},
  499--512.

\bibitem[\protect\citeauthoryear{Sterling and Shapiro}{Sterling and
  Shapiro}{1994}]{Sterling:Shapiro:94}
{\sc Sterling, L.} {\sc and} {\sc Shapiro, E.} 1994.
\newblock {\em The Art of Prolog\/}, 2nd ed.
\newblock MIT Press.

\bibitem[\protect\citeauthoryear{Swift, Warren, Sagonas, Freire, Rao, Cui,
  Johnson, de~Castro, Marques, Saha, Dawson, and Kifer}{Swift
  et~al\mbox{.}}{2022}]{xsb22}
{\sc Swift, T.}, {\sc Warren, D.~S.}, {\sc Sagonas, K.}, {\sc Freire, J.}, {\sc
  Rao, P.}, {\sc Cui, B.}, {\sc Johnson, E.}, {\sc de~Castro, L.}, {\sc
  Marques, R.~F.}, {\sc Saha, D.}, {\sc Dawson, S.}, {\sc and} {\sc Kifer, M.}
  2022.
\newblock {\em The XSB System Version 5.0,x}.
\newblock \url{http://xsb.sourceforge.net}. Latest release May 12, 2022.

\bibitem[\protect\citeauthoryear{Tamaki and Sato}{Tamaki and
  Sato}{1986}]{TamSat86}
{\sc Tamaki, H.} {\sc and} {\sc Sato, T.} 1986.
\newblock {OLD} resolution with tabulation.
\newblock In {\em Proceedings of the 3rd International Conference on Logic
  Programming}. Springer, 84--98.

\bibitem[\protect\citeauthoryear{Tekle and Liu}{Tekle and
  Liu}{2010}]{TekLiu10RuleQuery-PPDP}
{\sc Tekle, K.~T.} {\sc and} {\sc Liu, Y.~A.} 2010.
\newblock Precise complexity analysis for efficient {Datalog} queries.
\newblock In {\em Proceedings of the 12th International ACM SIGPLAN Symposium
  on Principles and Practice of Declarative Programming}. 35--44.
\newblock \url{https://doi.org/10.1145/1836089.1836094}.

\bibitem[\protect\citeauthoryear{Tekle and Liu}{Tekle and
  Liu}{2011}]{TekLiu11RuleQueryBeat-SIGMOD}
{\sc Tekle, K.~T.} {\sc and} {\sc Liu, Y.~A.} 2011.
\newblock More efficient {Datalog} queries: {S}ubsumptive tabling beats magic
  sets.
\newblock In {\em Proceedings of the 2011 ACM SIGMOD International Conference
  on Management of Data}. 661--672.
\newblock \url{http://doi.acm.org/10.1145/1989323.1989393}.

\bibitem[\protect\citeauthoryear{Tong, Lin, Liu, and Stoller}{Tong
  et~al\mbox{.}}{2023}]{alda23github}
{\sc Tong, Y.}, {\sc Lin, B.}, {\sc Liu, Y.~A.}, {\sc and} {\sc Stoller, S.~D.}
  2023.
\newblock {ALDA}.
\newblock \url{http://github.com/DistAlgo/alda}.
\newblock Accessed May 25, 2023.

\bibitem[\protect\citeauthoryear{Vennekens}{Vennekens}{2017}]{ven17idp-py}
{\sc Vennekens, J.} 2017.
\newblock Lowering the learning curve for declarative programming: {A Python
  API} for the {IDP} system.
\newblock In {\em Proceedings of 19th International Symposium on Practical
  Aspects of Declarative Languages}. Springer, 86--102.

\bibitem[\protect\citeauthoryear{Warren and Liu}{Warren and
  Liu}{2017}]{WarLiu17AppLP-arxiv}
{\sc Warren, D.~S.} {\sc and} {\sc Liu, Y.~A.} 2017.
\newblock {AppLP}: {A} dialogue on applications of logic programming.
\newblock {\em Computing Research Repository\/}~{\em arXiv:1704.02375 [cs.PL]}.
\newblock \url{http://arxiv.org/abs/1704.02375}.

\bibitem[\protect\citeauthoryear{Wright and Felleisen}{Wright and
  Felleisen}{1994}]{wright94syntactic}
{\sc Wright, A.~K.} {\sc and} {\sc Felleisen, M.} 1994.
\newblock A syntactic approach to type soundness.
\newblock {\em Information and Computation\/}~{\em 115}, 38--94.

\bibitem[\protect\citeauthoryear{Yang and Kifer}{Yang and
  Kifer}{2000}]{YanKif00flora}
{\sc Yang, G.} {\sc and} {\sc Kifer, M.} 2000.
\newblock {FLORA}: {I}mplementing an efficient {DOOD} system using a tabling
  logic engine.
\newblock In {\em Proceedings of the 1st International Conference on
  Computational Logic}. Springer, 1078--1093.
\newblock \url{https://doi.org/10.1007/3-540-44957-4_72}.

\bibitem[\protect\citeauthoryear{Zhou}{Zhou}{2016}]{Zhou16picat}
{\sc Zhou, N.-F.} 2016.
\newblock Programming in {Picat}.
\newblock In {\em Proceedings of the 10th International Symposium on Rule
  Technologies: Research, Tools, and Applications}. Springer, 3--18.

\end{thebibliography}
\bibliographystyle{\noticlp{alpha}\oops{plain}\iclp{acmtrans}}
}
}
\usebib
}

\oops{
\newpage
}
\appendix
\iclparxiv{


\newcommand{\firstalt}{~~}
\newcommand{\alt}{~~}

\newcommand{\negspc}{}

\newcommand{\pfn}{\rightharpoonup}
\newcommand{\bijection}{\stackrel{1-1}{\rightarrow}}
\newcommand{\union}{\cup}
\newcommand{\intersect}{\cap}
\newcommand{\UNION}{\bigcup}
\newcommand{\LAND}{\bigwedge}
\newcommand{\LOR}{\bigvee}
\newcommand{\Set}[1]{{\rm Set}(#1)}
\newcommand{\ra}{\rightarrow}
\newcommand{\myiff}{\Leftrightarrow}
\newcommand{\kwtt}[1]{\kw{\tt #1}}
\newcommand{\newaddr}{{\it newAddr}}

\newcommand{\Bool}{{\it Bool}}
\newcommand{\Int}{{\it Int}}
\newcommand{\Address}{\mathify{\it Address}}
\newcommand{\ProcessAddress}{{\it ProcessAddress}}
\newcommand{\NonProcessAddress}{{\it NonProcessAddress}}
\newcommand{\Val}{\mathify{\it Val}}
\newcommand{\Object}{{\it Object}}
\newcommand{\MsgQueue}{{\it MsgQueue}}
\newcommand{\ChannelStates}{{\it ChannelStates}}
\newcommand{\LocalHeap}{{\it LocalHeap}}
\newcommand{\Heap}{{\it Heap}}
\newcommand{\HeapType}{{\it HeapType}}
\newcommand{\State}{{\it State}}
\newcommand{\Program}{\mathify{\it Program}}
\newcommand{\Configuration}{\mathify{\it Configuration}}
\newcommand{\ProcessClass}{\mathify{\it ProcessClass}}
\newcommand{\Method}{\mathify{\it Method}}
\newcommand{\ReceiveDef}{\mathify{\it ReceiveDef}}
\newcommand{\ReceivePattern}{\mathify{\it ReceivePattern}}
\newcommand{\Pattern}{\mathify{\it Pattern}}
\newcommand{\InstanceVariable}{\mathify{\it InstanceVariable}}
\newcommand{\MethodName}{\mathify{\it MethodName}}
\newcommand{\Parameter}{\mathify{\it Parameter}}
\newcommand{\Expression}{\mathify{\it Expression}}
\newcommand{\Iterator}{\mathify{\it Iterator}}
\newcommand{\AnotherAwaitClause}{\mathify{\it AnotherAwaitClause}}
\newcommand{\Literal}{\mathify{\it Literal}}
\newcommand{\BooleanLiteral}{\Bool} 
\newcommand{\IntegerLiteral}{\Int} 
\newcommand{\UnaryOp}{\mathify{\it UnaryOp}}
\newcommand{\BinaryOp}{\mathify{\it BinaryOp}}
\newcommand{\TuplePattern}{\mathify{\it TuplePattern}}
\newcommand{\PatternElement}{\mathify{\it PatternElement}}
\newcommand{\ChannelOrder}{\mathify{\it ChannelOrder}}
\newcommand{\ChannelReliability}{\mathify{\it ChannelReliability}}
\newcommand{\EC}{\mathify{\it C}}
\newcommand{\LocalVariable}{\mathify{\it LocalVariable}}

\newcommand{\ClassName}{\mathify{\it ClassName}}
\newcommand{\Label}{\mathify{\it Label}}
\newcommand{\Field}{\mathify{\it Field}}
\newcommand{\Statement}{\mathify{\it Statement}}
\newcommand{\Tuple}{\mathify{\it Tuple}}

\newcommand{\commentMark}{/\,/}

\newcommand{\commentS}[1]{\mbox{\commentMark\ #1}}

\newcommand{\tuple}[1]{(#1)}
\newcommand{\ltuple}{(}
\newcommand{\rtuple}{)}
\newcommand{\seq}[1]{\langle#1\rangle}
\newcommand{\set}[1]{\{#1\}}

\newcommand{\state}{{\it state}}
\newcommand{\new}{{\it new}}
\newcommand{\emptySet}{\set{}}
\newcommand{\emptyfn}{\emptySet}
\newcommand{\emptyseq}{\seq{}}
\newcommand{\IF}{\mbox{if }}
\newcommand{\dom}{{\it dom}}
\newcommand{\range}{{\it range}}
\newcommand{\rest}{{\it rest}}
\newcommand{\length}{{\it length}}
\newcommand{\first}{{\it first}}
\newcommand{\bottom}{\mathord{\perp}}
\newcommand{\extends}{{\it extends}}
\newcommand{\methodDef}{{\it methodDef}}
\newcommand{\addrs}{{\it addrs}}
\newcommand{\subst}{{\it subst}}
\newcommand{\Init}{{\it Init}}
\newcommand{\tpli}{\hspace*{0.4em}}
\newcommand{\sci}{\hspace*{0.75em}}
\newcommand{\spce}{\hspace*{1em}}
\newcommand{\hole}{[\hspace*{0.1em}]}

\newcommand{\Class}{\mathify{\it Class}}
\newcommand{\Ruleset}{\mathify{\it Ruleset}}
\newcommand{\RulesetName}{\mathify{\it RulesetName}}
\newcommand{\Rule}{\mathify{\it Rule}}
\newcommand{\BasePredicate}{\mathify{\it BasePredicate}}
\newcommand{\DerivedPredicate}{\mathify{\it DerivedPredicate}}
\newcommand{\PredicateArg}{\mathify{\it PredicateArg}}
\newcommand{\KeywordArg}{\mathify{\it KeywordArg}}
\newcommand{\GlobalVariable}{\mathify{\it GlobalVariable}}
\newcommand{\LogicVariable}{\mathify{\it LogicVariable}}
\newcommand{\NLVariable}{\mathify{\it NonLocalVariable}}
\newcommand{\Query}{\mathify{\it Query}}
\newcommand{\Predicate}{\mathify{\it Predicate}}

\newcommand{\agv}{a_{\it gv}}
\newcommand{\classgv}{C_{\it gv}}
\newcommand{\pat}{{\it pat}}
\newcommand{\none}{\kwtt{None}}
\newcommand{\kwargs}{{\it kwargs}}
\newcommand{\evalrules}{{\it evalRules}}
\newcommand{\rulesetsg}{{\it glblRulesets}}
\newcommand{\rulesets}{{\it rulesets}}
\newcommand{\nlbase}{{\it nlBase}}
\newcommand{\nlderived}{{\it nlDerived}}
\newcommand{\derivedG}{{\it glblDerived}}
\newcommand{\rules}{{\it rules}}
\newcommand{\predicate}{{\it predicate}}
\newcommand{\pattern}{{\it pattern}}
\newcommand{\args}{{\it args}}
\newcommand{\maintain}{{\it maintain}}
\newcommand{\infupd}{{\it infUpdate}} 
\newcommand{\updatevar}{{\it updateVar}}
\newcommand{\legalassign}{{\it legalAssign}}
\newcommand{\result}{{\it result}}
\newcommand{\thetaht}{\theta_{T}}
\newcommand{\removevars}{{\it removeVars}}
\newcommand{\deref}{{\it deref}}
\newcommand{\allBaseAreSets}{{\it allBaseAreSets}}
\newcommand{\seminfer}{{\it infer}}
\newcommand{\rs}{{\it rs}}


\newcommand{\twocolonly}[1]{}






\section{Formal Semantics}
\label{app-formal}

\renewcommand{\topfraction}{0.99}	
\renewcommand{\bottomfraction}{0.99}	
\renewcommand{\textfraction}{0.01}	
\renewcommand{\dbltopfraction}{0.99}	
\renewcommand{\dblfloatpagefraction}{0.9}
\renewcommand{\floatpagefraction}{0.9}
\setcounter{topnumber}{2}
\setcounter{bottomnumber}{2}
\setcounter{totalnumber}{4}     
\setcounter{dbltopnumber}{2}

\newenvironment{ctabbing}
          {\begin{center}\begin{minipage}{\textwidth}\begin{tabbing}}
          {\end{tabbing}\end{minipage}\end{center}}



We give a complete abstract syntax and formal semantics for \iclp{our language}\noticlp{a language with objects and classes}.  The operational semantics is a reduction semantics with evaluation contexts \cite{wright94syntactic,serbanuta07rewriting}.  It builds on the standard least fixed-point semantics for Datalog~\cite{fitting2002fixpoint}
and the formal operational semantics for DistAlgo~\cite{Liu+17DistPL-TOPLAS}.
Relative to the latter, we removed the constructs specific to distributed algorithms, added an abstract syntax for rule sets and calls to \kwtt{infer}, added a transition rule for calls to \kwtt{infer}, extended the state with a stack that keeps track of rule sets whose results need to be maintained, extended several existing transition rules to perform automatic maintenance of the results of rule sets, and modified the semantics of existential quantifiers to bind the quantified variables to a witness when one exists.
The removed DistAlgo constructs can easily be restored;
we removed them simply to avoid repeating them.\noticlp{  
Compared to the core language in Section \ref{sec-formal}, the most significant additional features are: user-defined classes with inheritance; rule sets with class scope; allowing object fields as predicates in rule sets; set comprehension, quantification, and for-loop with tuple patterns; and support for tuple patterns in \kwtt{infer} statements.}

\subsection{Abstract syntax}
\label{sec:syntax}

The abstract syntax is defined in Figures \ref{fig:syntax1}--\ref{fig:syntax2}.  Tuples are immutable values, not mutable objects. Sets and sequences are mutable objects.  They are instances of the predefined classes \kwtt{set} and \kwtt{sequence}, respectively.  Methods of \kwtt{set} include \kwtt{add}, {\tt del}, \kwtt{contains}, \kwtt{size}, and \kwtt{any} (which returns an element of the set, if the set is non-empty, otherwise it returns \none).  Methods of \kwtt{sequence} include \kwtt{add} (which adds an element at the end of the sequence), \kwtt{contains}, and \kwtt{length}.  For brevity, among the standard arithmetic operations, we include only one representative operation in the abstract syntax and semantics; others are handled similarly.  All expressions are side-effect free.  Object creation, comprehension, and \kwtt{infer} are not expressions, because they all have the side-effect of creating one or more new objects. Semantically, the \kwtt{for} loop copies the contents of a (mutable) set or sequence into an (immutable) tuple before iterating over it, to ensure that changes to the set or sequence by the loop body do not affect the iteration.  \kwtt{whileSome} and \kwtt{ifSome} are similar to \kwtt{while} and \kwtt{if}, except that they always have an existential quantification as their condition, and they bind the variables in the pattern in the quantification to a witness, if one exists.  
We use some syntactic sugar in sample code, e.g., we use infix notation for some binary operators, such as \kwtt{is} and \kwtt{and}.

We refer to rule sets defined in global scope and class scope as ``global rule sets'' and ``class scope rule sets'', respectively.

Note that method parameters are not variables and cannot be assigned to, and that methods do not have local variables.  These choices simplify the semantics by eliminating the need for a call stack.  The only local variables are local variables of rule sets.  We refer to the other kinds of variables, namely global variables and instance variables, as non-local variables.  For brevity, we use ``variables'' without a qualifier to refer to non-local variables.



\begin{figure*}[htb]
\noticlp{\setstretch{0.97}}
\hspace*{1em}
\begin{ctabbing}
\Program\ ::= \Ruleset* \Class*\ \Statement\\
\Ruleset\ ::= \kwtt{rules} \RulesetName\ \Rule+\\
\Rule\ ::= \DerivedPredicate(\PredicateArg*)\ \kwtt{if} \BasePredicate(\PredicateArg*)*\\
\DerivedPredicate\ ::= \= \GlobalVariable\\
\> \kwtt{self}.\Field\\
\> \LocalVariable\\
\BasePredicate\ ::= \= \GlobalVariable[.\Field*]\\
\> \kwtt{self}.\Field+\\
\> \LocalVariable\\
\PredicateArg\ ::= \= \LogicVariable\\
\> \Literal \\
\Class\ ::= {\tt class} \ClassName\ [\kwtt{extends} \ClassName]\ : \Ruleset* \Method*\\
\Method\ ::= \=
 \kwtt{def} \MethodName{\tt (}\Parameter*{\tt )} \Statement\\
\> \kwtt{defun} \MethodName{\tt (}\Parameter*{\tt )} \Expression\\
\Statement\ ::= \= \NLVariable\ {\tt :=} \Expression\\
\> \NLVariable\ {\tt :=} \kwtt{new} \ClassName\\
\> \NLVariable\ {\tt := \{} \Expression\ {\tt :} \Iterator* {\tt |} \Expression\ {\tt \}}\\
\> \Statement\ {\tt ;} \Statement\\
\> \kwtt{if} \Expression\ {\tt :} \Statement\ {\tt \kw{else} :} \Statement\\
\> \kwtt{for} \Iterator\ {\tt :} \Statement\\
\> \kwtt{while} \Expression\ {\tt :} \Statement\\
\> \kwtt{ifSome} \Iterator\ {\tt |} \Expression\ {\tt :} \Statement\\
\> \kwtt{whileSome} \Iterator\ {\tt |} \Expression\ {\tt :} \Statement\\
\> \Expression{\tt .}\MethodName{\tt (}\Expression*{\tt )}\\
\> \NLVariable*\ := [\Expression{\tt .}]\kwtt{infer}(\= \Query*, \KeywordArg*,\\
\>\> \kwtt{rules}=\RulesetName)\\
\> \kwtt{skip}\\
\Expression\ ::= \=
\Literal\\
\> \Parameter\\
\> \NLVariable\\
\> \Tuple\\
\> \UnaryOp{\tt (}\Expression{\tt )}\\
\> \BinaryOp{\tt (}\Expression{\tt ,}\Expression{\tt )}\\
\> \kwtt{isinstance}{\tt (}\Expression,\ClassName{\tt )}\\
\> \kwtt{and}{\tt (}\Expression,\Expression{\tt )} \spce \= \commentMark\ conjunction (short-circuiting)\\
\> \kwtt{or}{\tt (}\Expression,\Expression{\tt )}        \> \commentMark\ disjunction (short-circuiting)\\
\> \kwtt{each} \Iterator\ {\tt |} \Expression\\
\> \kwtt{some} \Iterator\ {\tt |} \Expression\\
\> \Expression{\tt .}\MethodName{\tt (}\Expression*{\tt )}
\end{ctabbing}
\caption{Abstract syntax, Part 1.}
\label{fig:syntax1}
\end{figure*}

\begin{figure}[htb]
\begin{ctabbing}
\NLVariable\ := \= \GlobalVariable\\
\> \InstanceVariable\\
\InstanceVariable\ ::= \Expression.\Field\\
\Literal\ ::= \=
  \kwtt{None}\\
\> \BooleanLiteral\\
\> \IntegerLiteral\\
\> ...\\
\BooleanLiteral\ ::= \=
   \kwtt{True}\\
\> \kwtt{False}\\
\IntegerLiteral\ ::= ...\\
\Iterator\ ::= \Pattern\ \kwtt{in} \Expression\\
\Pattern\ ::= \=
   \NLVariable\\
\> \TuplePattern\\
\TuplePattern\ ::= {\tt (}\PatternElement*{\tt )}\\
\PatternElement\ ::= \=
   \Expression\\
\> \_\\
\> {\tt =}\NLVariable\\
\Query\ := \=\Predicate [\TuplePattern]\\
\KeywordArg\ ::= \LocalVariable\ = \Expression\\
\Tuple\ ::= {\tt (}\Expression*{\tt )}\\
\UnaryOp\ ::= \=
   \kwtt{not}    \hspace*{3em} \= \commentMark\ Boolean negation\\
\> \kwtt{isTuple}           \> \commentMark\ test whether a value is a tuple\\
\> \kwtt{len}               \> \commentMark\ length of a tuple\\
\BinaryOp\ ::= \=
   \kwtt{is}           \> \commentMark\ identity-based equality\\
\> \kwtt{plus}         \> \commentMark\ sum\\
\> \kwtt{select}       \> \commentMark\ \kwtt{select}($t$,$i$) returns the $i$'th\twocolonly{\\ \>\> \commentMark } component of tuple $t$
\end{ctabbing}
  \caption{Abstract syntax, Part 2. Ellipses (``...'') are used for common syntactic categories whose details are unimportant.  Details of the identifiers allowed for non-terminals \RulesetName, \GlobalVariable, \Field, \LocalVariable, \LogicVariable, \ClassName, \MethodName, and \Parameter\ are also unimportant and hence unspecified, except that \ClassName\ must include \kwtt{set} and \kwtt{sequence}, and \Parameter\ must include \kwtt{self}.
  }
\label{fig:syntax2}
\end{figure}

\mypar{Notation in the grammar}
%
%
A symbol in the grammar is a terminal symbol if it is in typewriter font or 
is a non-terminal symbol if it is in italics.
In each production, alternatives are separated by a linebreak.
Square brackets enclose optional clauses.
{\tt *} after a non-terminal means ``0 or more occurrences''.
{\tt +} after a non-terminal means ``1 or more occurrences''.

\mypar{Well-formedness requirements on programs}
In global rule sets, predicates cannot contain \kwtt{self}.
In class scope rule sets, derived predicates cannot be global variables.
Each global variable appears as a derived predicate in at most one rule set in the program.  
In each class, for each field $f$, \kwtt{self}.$f$ appears as a derived predicate in at most one rule set in that class.
In each rule in each rule set, each logic variable that appears in the conclusion appears in a hypothesis.  Within each rule set, all uses of the same predicate have the same number of arguments.


In assignments with a call to \co{\INFER} on the right side of the assignment, the number of variables on the left of the assignment equals the number of queries, the predicates queried are derived predicates of the rule set used, and local variables in keyword arguments are base predicates of the rule set used.

Invocations of methods defined using \kwtt{def} appear only as statements.  Invocations of methods defined using \kwtt{defun} appear only as expressions; we also refer to these methods as ``functions''.  The program does not contain definitions of classes named \kwtt{set} and \kwtt{sequence}.

Class names are unique; in other words, each class name is defined at most once.  
Method names are unique within the scope of each class.
Rule set names are unique within each scope.

\subsubsection{Constructs whose semantics is given by translation}
\label{sec:translation}


\mypar{Notation}
A (partial) function is represented as a set of mappings $x \mapsto y$.  We represent substitutions as functions from parameters and variables to expressions.  $t\, \theta$ denotes the result of applying substitution $\theta$ to $t$.

\mypar{Class scope rule set names}  The program is transformed so that rule set names are unique across all scopes.  A straightforward way to do this is to prefix the name of every class scope rule set with the name of the enclosing class.

\mypar{Global variables} Global variables are replaced with instance variables, and global rule sets are transformed to have the same form as class scope rule sets, by the following transformations.  Choose an address $\agv$ whose fields will be used to represent global variables.  Everywhere except in global rule sets and in queries in calls to \kwtt{infer} on global rule sets, replace each global variable $x$ with $\agv.x$.  Introduce a class name $\classgv$, put all global rule sets into this class, and replace each global variable $x$ with \kwtt{self}.$x$ in those rule sets and in queries in calls to \kwtt{infer} on those rule sets.  
Calls to \kwtt{infer} on those rule sets are also transformed by prefixing $\agv{\tt .}$ to the call, i.e., $\kwtt{infer}(\cdots)$ is replaced with $\agv{\tt .}\kwtt{infer}{\tt (}\cdots{\tt )}$, so all calls to \kwtt{infer} have a target object.  The initial state of the program is defined so that an object of type $\classgv$ is at address $\agv$.  These transformations simplify the transition rules related to inference, by allowing global rule sets and class scope rule sets to be handled in a uniform way.

\mypar{Boolean operators} The Boolean operators \kwtt{and} and \kwtt{each} are eliminated as follows: {\tt $e_1$ \kw{and} $e_2$} is replaced with {\tt \kw{not}(\kw{not}($e_1$) \kw{or} \kw{not}($e_2$))}, and {\tt \kw{each} {\it iter} | $e$} is replaced with {\tt \kw{not}(\kw{some} {\it iter} | \kw{not}($e$))}.

\mypar{Non-variable expressions in tuple patterns} Non-variable expressions in tuple patterns are replaced with variables prefixed by ``{\tt =}''.  Specifically, for each expression $e$ in a tuple pattern that is not a variable, a variable prefixed with ``{\tt =}'', or wildcard, an assignment $v$~{\tt :=}~$e$ to a fresh variable $v$ is inserted before the statement that contains the tuple pattern, and $e$ is replaced with {\tt =}$v$ in the tuple pattern.

\mypar{Wildcards} Wildcards are eliminated from tuple patterns in \kwtt{for} loops, comprehensions, and quantifications (i.e., everywhere except as $\Query$ in \kwtt{infer}) by replacing each wildcard with a fresh variable.


\mypar{Tuple patterns in \kwtt{infer} statements} \hspace{0pt}\kwtt{infer} statements are transformed to eliminate tuple patterns in queries.  After transformation, each query is simply the name of a predicate. 
 Consider the statement {\tt 
$x_1,\ldots,x_n$ := $[e.]$\kw{infer}($p_1(\pat_1),$ $\ldots,$ $p_n(pat_n), \kwargs,$ \kw{rules}=$rs$)}.
Let $x_{i,1},$ $\ldots,$ $x_{i,k_i}$ be the components of $\pat_i$, in order and without repetitions, that are variables not prefixed by ``{\tt =}''.  Let $y_1,\ldots,y_n$ be fresh variables.  The above statement is transformed to: 
\begin{alltt}
\(y\sb{1},\ldots,y\sb{n}\) := \([e.]\)\kw{infer}(\(p\sb{1},\ldots,p\sb{n}, \kwargs,\) \kw{rules}=\(rs\))
\(x\sb{1}\) := \{ (\(x\sb{1,1},\ldots,x\sb{1,k\sb{1}}\)) : \(\pat\sb{1}\) \kw{in} \(y\sb{1}\) | \kw{True} \}
\(\ldots\)
\(x\sb{n}\) := \{ (\(x\sb{n,1},\ldots,x\sb{n,k\sb{n}}\)) : \(\pat\sb{n}\) \kw{in} \(y\sb{n}\) | \kw{True} \}
\end{alltt}

\mypar{\kwtt{ifSome} statements} \hspace{0pt}\kwtt{ifSome} is statically eliminated as follows.  Consider the statement {\tt \kw{ifSome} \pat\ \kw{in} $e$ | $b$ :~$s$}.  Let $i_1,\ldots,i_k$ be indices, in order of appearance from left to right, of elements of \pat\ that are variables not prefixed by ``{\tt =}''.  Let $x_{i_1},\ldots,x_{i_k}$ be those variables.  Let {\tt foundOne} and $x'_{i_1},\ldots,x'_{i_k}$ be fresh variables.  Let substitution $\theta$ be $[x_{i_1}\mapsto x'_{i_1}, \ldots, x_{i_k}\mapsto x'_{i_k}]$.  Let $\pat' = \pat\,\theta$ and $b' = b\,\theta$.
The above \kwtt{ifSome} statement is transformed to:
\begin{alltt}
foundOne := \kw{False}
\kw{for} \(pat'\) \kw{in} \(e\):
  \kw{if} \(b'\) \kw{and} \kw{not} foundOne:
    \(x\sb{i\sb{1}} := x'\sb{i\sb{1}}\)
    \(\ldots\)
    \(x\sb{i\sb{k}} := x'\sb{i\sb{k}}\)
    \(s\)
    foundOne := \kw{True}
\end{alltt}

\mypar{\kwtt{whileSome} statements} \hspace{0pt}\kwtt{whileSome} is statically eliminated as follows.  Consider the statement {\tt \kw{whileSome} \pat\ \kw{in} $e$ | $b$ :~$s$}.  Using the same definitions as in the previous item, this statement is transformed to:
\begin{alltt}
foundOne := \kw{True}
\kw{while} foundOne:
  foundOne := \kw{False}
  \kw{for} \(pat'\) \kw{in} \(e\):
    \kw{if} \(b'\) \kw{and} \kw{not} foundOne: 
      \(x\sb{i\sb{1}} := x'\sb{i\sb{1}}\)
      \(\ldots\)
      \(x\sb{i\sb{k}} := x'\sb{i\sb{k}}\)
      \(s\)
      foundOne := \kw{True}
\end{alltt}


\mypar{Comprehensions}  First, comprehensions are transformed to eliminate the use of variables prefixed with ``{\tt =}''.  Specifically, for a variable $x$ prefixed with ``{\tt =}'' in a comprehension, replace occurrences of {\tt =}{\it x} in the comprehension with occurrences of a fresh variable $y$, and add the conjunct {\tt $y$ is $x$} to the Boolean condition.  Second,
all comprehensions are statically eliminated as follows.  The comprehension {\tt $x$ := \{ $e$ | $pat_1$ \kw{in} $e_1$, $\ldots$, $pat_n$ \kw{in} $e_n$ | $b$ \}} is replaced with
\begin{alltt}
\(x\) := \kw{new} \kw{set}
\kw{for} \(pat\sb{1}\) \kw{in} \(e\sb{1}\):
  ...
    \kw{for} \(pat\sb{n}\) \kw{in} \(e\sb{n}\):
      \kw{if} \(b\):
        \(x\).\kw{add}(\(e\))
\end{alltt}

\mypar{Tuple patterns in iterators} Iterators containing tuple patterns are rewritten as iterators without tuple patterns.


Consider the existential quantification {\tt \kw{some} ($e_1,
    \ldots, e_n$) \kw{in} $e$ | $b$}.  Let $x$ be a fresh variable.  Let
  $\theta$ be the substitution that replaces $e_i$ with {\tt
    select($x$,$i$)} for each $i$ such that $e_i$ is a variable not
  prefixed with ``{\tt =}''.  Let $\{j_1,\ldots,j_m\}$ contain the indices
  of the constants and the variables prefixed with ``{\tt =}'' in {\tt
    ($e_1, \ldots ,e_n$)}.  Let $\bar e_j$ denote $e_j$ after removing the
  ``{\tt =}'' prefix, if any.  The quantification is rewritten as {\tt \kw{some}
    $x$ \kw{in} $e$ | \kw{isTuple}($x$) \kwtt{and} \kw{len}($x$) \kw{is} $n$ and (\kw{select}($x$,$j_1$),
    $\ldots$, \kw{select}($x$,$j_m$)) \kw{is} ($\bar e_{j_1}$, $\ldots$, $\bar
    e_{j_m}$) \kw{and} $b\,\theta$}.


Consider the loop {\tt \kw{for} ($e_1,\ldots,e_n$) \kw{in} $e$ :~$s$}.  Let $x$ and
 $S$ be fresh variables.  Let $\{i_1,\ldots,i_k\}$ contain the indices
  in {\tt ($e_1, \ldots ,e_n$)} of variables not prefixed with ``{\tt =}''.
  Let $\{j_1,\ldots,j_m\}$ be as in the previous paragraph.
  Let $\bar e_j$ denote $e_j$ after removing the ``{\tt =}'' prefix, if any.  Note
  that $e$ may evaluate to a set or sequence, and duplicate bindings for the
  tuple of variables $(e_{i_1},\ldots,e_{i_k})$ are filtered out if $e$ evaluates
  to a set but not if $e$ evaluates to a sequence.  The loop is rewritten as the code in
  Figure \ref{fig:elim-tuple-from-for-loop}.
 
\begin{figure}[htb]
\begin{alltt}
    \(S\) := \(e\)
    \kw{if} \kw{isinstance}(\(S\),\kw{set}):
      \(S\) := \{ \(x\) : \(x\) in \(S\) | \kw{isTuple}(\(x\)) \kw{and} \kw{len}(\(x\)) \kw{is} \(n\)
          \kw{and} (\kw{select}(\(x\),\(j\sb{1}\)), \(\ldots\), \kw{select}(\(x\),\(j\sb{m}\)))
             \kw{is} (\(\bar{e}\sb{j\sb{1}}\), \(\ldots\), \(\bar{e}\sb{j\sb{m}}\)) \}
      \kw{for} \(x\) \kw{in} \(S\): 
        \(e\sb{i\sb{1}} := \kwtt{select}(x,i\sb{1})\)
        \(\ldots\)
        \(e\sb{i\sb{k}} := \kwtt{select}(x,i\sb{k})\)
        \(s\)
    \kw{else}: {\rm \commentMark \(S\) is a sequence}
      \kw{for} \(x\) \kw{in} \(S\): 
        \kw{if} (\kw{isTuple}(\(x\)) \kw{and} \kw{len}(\(x\)) \kw{is} \(n\) 
            \kw{and} (\kw{select}(\(x\),\(j\sb{1}\)), \(\ldots\), \kw{select}(\(x\),\(j\sb{m}\)))
               \kw{is} (\(\bar{e}\sb{j\sb{1}}\), \(\ldots\), \(\bar{e}\sb{j\sb{m}}\)): 
          \(e\sb{i\sb{1}} := \kwtt{select}(x,i\sb{1})\)
          \(\ldots\)
          \(e\sb{i\sb{k}} := \kwtt{select}(x,i\sb{k})\)
          \(s\)
        \kw{else}: 
          \kw{skip}
\end{alltt}
\caption{Translation of \kwtt{for} loop to eliminate tuple pattern.}
\label{fig:elim-tuple-from-for-loop}
\end{figure}

\subsection{Semantic domains}
\label{sec:domains}

The semantic domains are defined in Figure \ref{fig:domains}, using the following notation.  $D^*$ is the set of finite sequences of values from domain $D$.  $\Set{D}$ is the set of finite sets of values from domain $D$. $D_1 \ra D_2$ and $D_1 \pfn D_2$ are the sets of (total) functions and partial functions, respectively, from $D_1$ to $D_2$.  $\dom(f)$ and $\range(f)$ are the domain and range, respectively, of a partial function $f$, i.e., $\dom(f) = \set{x \;|\; \exists y : x \mapsto y \in f}$ and $\range(f) = \set{y \;|\; \exists x : x \mapsto y \in f}$.

In a state $\tuple{s,h,ht}$, $s$ is the statement to be executed, $h$ is the heap that maps an address to the object at that address, and $ht$ is the heap type map that maps an address to the type of the object on the heap at that address.\noticlp{  Note that the environment used in the core semantics in Section \ref{sec-formal} to store values of global variables is not needed here, because global variables have been replaced with object fields whose values are stored on the heap.}

\begin{figure}[htb]
\begin{eqnarray*}
  \Bool \negspc&=&\negspc \set{\kwtt{True}, \kwtt{False}}\\
  \Int \negspc&=&\negspc ... \\
  \Address \negspc&=&\negspc ...\\
  \Tuple \negspc&=&\negspc \Val^*\\
  \Val \negspc&=&\negspc \Bool \union \Int \union \Address \union \Tuple \union \{\none\}\\
  \Object \negspc&=&\negspc (\Field \pfn \Val) \union \Set{\Val} \union \Val^*\\
  \HeapType \negspc&=&\negspc \Address \pfn \ClassName\\
  \Heap \negspc&=&\negspc \Address \pfn \Object\\
  \State \negspc&=&\negspc \Statement \times \Heap \times \HeapType
\end{eqnarray*}
\caption{Semantic domains.  Ellipses are used for semantic domains of primitive values whose details are standard or unimportant.}
  \label{fig:domains}
\end{figure}

\subsection{Extended abstract syntax}
\label{sec:extended-syntax}

Section \ref{sec:syntax} defines the abstract syntax of programs that can be written by the user.  We extend the abstract syntax to include additional forms into which programs may evolve during evaluation.  The new productions appear below.   The statement {\tt \kw{for} $v$ \kw{inTuple} $t$:~$s$} iterates over the elements of tuple $t$, in the obvious way.

\begin{ctabbing}
\Expression\ ::= \= {\it Address}\\
\> {\it Address}.\Field\\
\\
\Statement\ ::= \kwtt{for} {\it Variable} \kwtt{inTuple} \Tuple: \Statement
\end{ctabbing}

\subsection{Evaluation contexts}

Evaluation contexts, also called reduction contexts, are used to identify
the next part of an expression or statement to be evaluated.  An evaluation
context is an expression or statement with a hole, denoted \hole, in
place of the next sub-expression or sub-statement to be evaluated.  Evaluation contexts are defined in Figure \ref{fig:eval-context}.  Note that square brackets enclosing a clause indicate that the clause is optional; this is unrelated to the notation \hole\ for the hole.  

For example, the definition of evaluation contexts for method calls (lines 3--4 of Figure \ref{fig:eval-context}) says that the expression denoting the target object is evaluated first to obtain an address (if the expression isn't already an address); then, the arguments are evaluated from left to right.  The left-to-right order holds because an argument can be evaluated only if the arguments to its left are values, as opposed to more complicated unevaluated expressions.  The definition of evaluation contexts for \kwtt{infer} implies that the expressions for the targets of the assignment are evaluated from left to right; then the expression for the target object, if any (i.e., if the call is for a rule set with class scope), is evaluated; and then the argument expressions are evaluated from left to right.
 


\begin{figure}[htb]
\begin{ctabbing}
\EC\ ::= \=
 \hole\\
\> (\Val*, \EC, \Expression*)\\
\> \EC.\MethodName(\Expression*)\\
\> \Address.\MethodName(\Val*, \EC, \Expression*)\\
\> \UnaryOp(\EC)\\
\> \BinaryOp(\EC, \Expression)\\
\> \BinaryOp(\Val, \EC)\\
\> \kwtt{isinstance}(\EC, \ClassName)\\
\> \kwtt{or}(\EC, \Expression)\\
\> \kwtt{some} \Pattern\ \kwtt{in} \EC\ {\tt |} \Expression\\
\> \EC.\Field\ {\tt :=} \Expression\\
\> \EC.\Field\ {\tt :=} \kwtt{new} \ClassName\\
\> \Address.\Field\ {\tt :=} \EC\\
\> \EC\ ; \Statement\\
\> \kwtt{if} \EC{\tt :} \Statement\ {\tt \kw{else}:} \Statement\\
\> \kwtt{for} \InstanceVariable\ \kwtt{in} {\it \EC}{\tt :} \Statement\\
\> \kwtt{for} \InstanceVariable\ \kwtt{inTuple} \Tuple{\tt :} \EC\\
\> (\Address.\Field)*, \EC.\Field, (\Expression.\Field)* {\tt :=}\\
\> \hspace*{1em}[\Expression.]{\tt \kw{infer}(}\=\Query*, \KeywordArg*,\\
\>\>{\tt \kw{rules}=\RulesetName)}\\
\> (\Address.\Field)* {\tt :=} \EC.{\tt \kw{infer}(}\=\Query*, \KeywordArg*,\\
\>\>{\tt \kw{rules}=\RulesetName)}\\
\> (\Address.\Field)* {\tt :=}\\
\> \hspace*{1em}[\Address.]{\tt \kw{infer}(}\Query*, (\Parameter=\Val)*,\\
\> \hspace*{1em}{\tt \Parameter=\EC, \KeywordArg{}*, \kw{rules}=\RulesetName)}
\end{ctabbing}
   \caption{Evaluation contexts for expressions and statements.}
   \label{fig:eval-context}
 \end{figure}

\subsection{Transition relations}
\label{sec:transition}

The transition relation for expressions has the form $h,ht \vdash e \ra e'$, where $e$ and $e'$ are expressions, $h\in\Heap$, and $ht\in\HeapType$. The transition relation for statements has the form $\state \ra \state'$ where $\state\in\State$ and $\state'\in\State$.  

Both transition relations, and some of the auxiliary functions defined below, are implicitly parameterized by the program, which is needed to look up method definitions, rule set definitions, etc.  The transition relation for expressions is defined in Figure \ref{fig:transition-expr}.  The transition relation for statements is defined in Figures \ref{fig:transition-one}--\ref{fig:transition-two}, using auxiliary functions defined in Figure \ref{fig:maintain}.  The context rules for expressions and statements at the top of Figure \ref{fig:transition-one} allow the expression or statement in the evaluation context's hole to take a transition, while the rest of the program, denoted by $C$, is carried along unchanged.

\begin{figure}[tbp]
\begin{displaymath}
\begin{array}{@{}l@{}}
\commentS{field access}\\
h,ht \vdash a.f \ra h(a)(f) \spce \IF ht(a) \not\in\set{\kwtt{set},\kwtt{seq}} \land f \in \dom(h(a))\\
\\
\commentS{invoke function in user-defined class}\\
h,ht \vdash a{\tt .}m(v_1,\ldots,v_n) \ra e[\kwtt{self}\mapsto a, x_1\mapsto v_1, \ldots, x_n\mapsto v_n]\\
\sci \IF \methodDef(ht(a), m, \kwtt{defun}~m(x_1,\ldots,x_n)~e)\\
\\
\commentS{invoke function in pre-defined class (example)}\\
h,ht \vdash a{\tt .\kw{any}()} \ra v
\spce \IF ht(a)=\kwtt{set} \land v \in h(a)\\
h,ht \vdash a{\tt .\kw{any}()} \ra \kwtt{None}
\spce \IF ht(a)=\kwtt{set} \land h(a)=\emptyset\\
\\
\commentS{unary operations}\\
h,ht \vdash {\tt \kw{not}(\kw{True})} \ra \kwtt{False}\\
h,ht \vdash {\tt \kw{not}(\kw{False})} \ra \kwtt{True}\\
h,ht \vdash \kwtt{isTuple}{\tt (}v{\tt )} \ra \kwtt{True} \spce 
\mbox{if $v$ is a tuple}\\
h,ht \vdash \kwtt{isTuple}{\tt (}v{\tt )} \ra \kwtt{False} \spce 
\mbox{if $v$ is not a tuple}\\
h,ht \vdash \kwtt{len}{\tt (}v{\tt )} \ra n  \spce 
\mbox{if $v$ is a tuple with $n$ components}\\
\\
\commentS{binary operations}\\
h,ht \vdash {\tt \kw{is}(}v_1,v_2{\tt )} \ra \kwtt{True} \\
\sci \mbox{if $v_1$ and $v_2$ are the same (identical) value}\\
\\
h,ht \vdash {\tt \kw{plus}(}v_1,v_2{\tt )} \ra v_3 \\
\sci \IF v_1\in\Int \land v_2\in\Int \land v_3 = v_1 + v_2\\
\\
h,ht \vdash {\tt \kw{select}(}v_1,v_2{\tt )} \ra v_3\\
\sci \IF v_2\in\Int \land v_2>0 \land
\mbox{($v_1$ is a tuple with length at least $v_2$)}\\
\sci {} \land \mbox{($v_3$ is the $v_2$'th component of $v_1$)}\\
\\
\commentS{isinstance}\\
h,ht \vdash {\tt \kw{isinstance}(}a, c{\tt )} \ra \kwtt{True} \spce \IF ht(a)=c\\
h,ht \vdash {\tt \kw{isinstance}(}a, c{\tt )} \ra \kwtt{False} \spce \IF ht(a)\ne c\\
\\
\commentS{disjunction}\\
h,ht \vdash {\tt \kw{or}(\kw{True}}, e{\tt )} \ra \kwtt{True}\\
h,ht \vdash {\tt \kw{or}(\kw{False}}, e{\tt )} \ra e\\
\\
\commentS{existential quantification}\\
h,ht \vdash \kwtt{some}~x~\kwtt{in}~a~|~e ~~\ra~~ e[x\mapsto v_1]~{\tt \kw{or} }~\cdots~\kwtt{or}~e[x\mapsto v_n]\\
\sci \IF (ht(a)=\kwtt{sequence} \land h(a)=\seq{v_1,\ldots,v_n}) \\
\sci {} \lor (ht(a)=\kwtt{set} \land \seq{v_1,\ldots,v_n}~\mbox{is a linearization of}~h(a))
\end{array}
\end{displaymath}
  \caption{Transition relation for expressions.}
  \label{fig:transition-expr}
\end{figure}

\begin{figure}[htb]
\begin{displaymath}
\begin{array}{@{}l@{}}
\deref(h,a,F) = 
\begin{array}[t]{@{}l@{}}
  \mbox{if } a \not\in \dom(h) \mbox{ then } \bottom\\
  \mbox{elif } \length(F)=1 \mbox{ then } 
  (\mbox{if } F \in \dom(h(a)) \mbox{ then } h(a)(F) \mbox{ else } \bottom)\\
  \mbox{else } \first(F) \in \dom(h(a)) \mbox{ then } \deref(h, h(a)(\first(F)),\rest(path)) \mbox{ else } \bottom\\
\end{array}\\
\\
\allBaseAreSets(h,ht) = \\
\sci
\begin{array}[t]{@{}l@{}}
\forall a \in \dom(h), \rs \in \rulesets(ht(a)), \kwtt{self}.F \in \nlbase(\rules(\rs)) :\\
 \deref(h, a, F) = \bottom \lor
 (\deref(h, a, F) \in \Address \land ht(\deref(h, a, F)) = \kwtt{set})
\end{array}\\
\\
\updatevar(h,\rs,a.f,S) = 
\begin{array}[t]{@{}l@{}}
 \mbox{if } h(a)(f) \in \Address \mbox{ then } \set{h(a)(f) \mapsto S} \\
 \mbox{else } \set{a \mapsto h(a)[f \mapsto \newaddr(\rs,a.f,h)], \newaddr(\rs,a.f,h) \mapsto S}
\end{array}\\
\\
\infupd(h,\rs,a,\args) = \\
\sci \begin{array}{@{}l@{}}
\mbox{let }
\begin{array}[t]{@{}l@{}}
  {\it facts}_B = \set{a.F(v) : \kwtt{self}.F \in \nlbase(\rules(\rs)) \land v \in \deref(h,a,F)}\\
  {\it facts}_L = \set{p(v) : p \in \dom(\args) \land v \in h(\args(p))}\\
  \result = \evalrules(\rules(rs)[\kwtt{self}\mapsto a] \union {\it facts}_B \union {\it facts}_L)\\
  \theta = \UNION_{\kwtt{self}.f\in \nlderived(\rules(\rs))} \updatevar(h,\rs,a.f,\result(a.f))
\end{array}\\
\mbox{in } \tuple{\theta, \result}
\end{array}\\
\\
\maintain(h,ht) = 
\begin{array}[t]{@{}l@{}}
  \mbox{let } 
    \begin{array}[t]{@{}l@{}}
    \theta = \UNION_{a \in \dom(h), \rs \in \rulesets(ht(a))} \pi_1(\infupd(h,\rs,a,\emptyfn))\\
    \thetaht = \set{a \mapsto \kwtt{set} \;|\; a \in \dom(\theta) \land \theta(a) \subseteq \Val})\\
    \tuple{h', ht'} = \tuple{h, ht} \sqcup \tuple{\theta, \thetaht}
  \end{array}\\
  \mbox{in if } (h',ht')=(h,ht) \mbox{ then } (h,ht) \mbox{ else } \maintain(h',ht')
\end{array}
\end{array}
\end{displaymath}
\caption{Definitions of auxiliary functions related to inference.}
\label{fig:maintain}
\end{figure}

\mypar{Notation}
In the transition rules, $a$ matches an address, and $v$ matches a value (i.e., an element of $\Val$).

$f \union g$ is the union of functions $f$ and $g$ with disjoint domains.  For any functions $f$ and $g$, $f \sqcup g = \set{x \mapsto f(x) \;|\; x \in \dom(f) \setminus \dom(g)} \union g$.  For a function $f$, $f[x \mapsto y] = f \sqcup \set{x\mapsto y}$.  When a function $\theta$ is intended to be used to compute an updated version $f \sqcup \theta$ of a function $f$, we refer to $\theta$ as an ``update'' to $f$.

  
Sequences are denoted with angle brackets, e.g., $\seq{0,1,2} \in \Int^*$.  $s@t$ is the concatenation of sequences $s$ and $t$. $\first(s)$ is the first element of sequence $s$.  $\rest(s)$ is the sequence obtained by removing the first element of $s$.  $\length(s)$ is the length of sequence $s$. 

\mypar{Auxiliary definitions}
$\new(c)$ returns a new instance of class $c$, for $c\in\ClassName$.  When $c$ is the name of user-defined class, $\new(c)$ returns an empty set representing the empty function.
\begin{displaymath}
  \begin{array}{@{}l@{}}
    \new(c) = \mbox{if } c=\kwtt{sequence} \mbox{ then } \emptyseq \mbox{ else } \emptyfn
  \end{array}
\end{displaymath}


$\legalassign(ht,a,f)$ holds if assigning to field $f$ of the object with address $a$ is legal, in the sense that $a$ refers to an object with fields (not an instance of a pre-defined class without fields), and $a.f$ is not a derived predicate of any rule set.  
$\legalassign(ht,a,f) = ht(a) \not\in\set{\kwtt{set}, \kwtt{sequence}} \land
((a = \agv \land \agv.f \not\in\derivedG) \lor 
(a \ne \agv \land {\tt \kw{self}.f} \not\in\nlderived(ht(a))))$.

$\methodDef(c, m, {\it def})$ holds iff $c$ is a user-defined class and either (1) $c$ defines  method $m$, and {\it def} is the definition of $m$ in $c$, or (2) $c$ does not define $m$, and {\it def} is the definition of $m$ in the nearest ancestor of $c$ in the inheritance hierarchy that defines $m$.

$\rulesetsg$ is the set of names of global rule sets in the program.  $\rulesets(c)$ is the set of names of rule sets defined in class $c$ in the program.  By definition, $\rulesets(\classgv)=\rulesetsg$, and for convenience, 
we also define $\rulesets(\kwtt{set})=\emptyset$ and $\rulesets(\kwtt{seq})=\emptyset$. 
For any rule set name $\rs$ in the program, $\rules(\rs)$ is the set of rules in that rule set (recall from Section \ref{sec:translation} that rule set names have been transformed to be unique across all scopes).

For a set of rules $R$, $\nlbase(R)$ and $\nlderived(R)$ are the sets of non-local base predicates and non-local derived predicates, respectively, in $R$.  For $c\in\ClassName$, $\nlderived(c)$ is the set of non-local derived predicates in rule sets defined in class $c$.  $\derivedG$ is the set of global variables that are derived predicates in any rule set.

For a derived predicate $p$ of rule set $\rs$, and heap $h$, $\newaddr(\rs,p,h)$ selects a fresh address for $p$ of $\rs$; specifically, it returns an address that is not in $\dom(h)$ and is different from $\newaddr(\rs',p',h)$ whenever $\rs \ne \rs' \lor p \ne p'$.  Using function $\newaddr$ to select fresh addresses, instead of selecting them non-deterministically, is inessential but simplifies the definitions of auxiliary functions related to inference in Figure~\ref{fig:maintain} and the transition rule for \kwtt{infer} in Figure~\ref{fig:transition-two}.

The following five auxiliary functions and relation are defined in Figure \ref{fig:maintain}.

$\deref(h,a,F)$ returns the value obtained by starting at address $a$ in heap $h$ and dereferencing the sequence $F$ of one or more fields.  If $a$ is not an address in $\dom(h)$, or if a field in $F$ is not in the domain of the appropriate object, then $\deref$ returns $\bottom$. 
  
$\allBaseAreSets(h,ht)$ returns true if, in heap $h$ with heap type map $ht$, for each rule set, for each non-local base predicate of the rule set, either it is uninitialized (indicated by $\deref$ returning $\bottom$) or its value is a set.

$\updatevar(h,\rs,a.f,S)$ returns an update to the heap that makes variable $a.f$ to refer to a set with content $S$.  If the value of $a.f$ is already an address $a'$, a set with content $S$ is stored at $a'$, otherwise $a.f$ is assigned a fresh address $a'$, and a set with content $S$ is stored at $a'$.

$\infupd(h,\rs,a,\args)$ computes an update expressing the result of inference for rule set $\rs$ instantiated with $\kwtt{self}\mapsto a$, with heap $h$ and using $\args$ to obtain values for local variables of $\ra$.  $\infupd$ returns a pair containing the update to apply to the heap $h$ and a function $\result$ that maps each defined derived predicate in the rule set to its value; a derived predicate is undefined after this inference if it depends on a local variable that is a base predicate whose value is not provided by $\args$.  For explicit calls to \kwtt{infer}, $\args$ contains values provided by keyword arguments; for automatic maintenance, $\args$ is the empty function.  Note that the condition $v \in \deref(h,a,F)$ is false if $\deref(h,a,F)$ is $\bottom$; this has the effect that uninitialized base predicates are equivalent to empty sets.  $\infupd$ uses the auxiliary function $\evalrules(R)$, which evaluates the set of rules $R$ and returns a function from the set of predicates that appear in the rules to their meanings, represented as sets of tuples.

$\maintain(\theta,\thetaht,h,ht)$ returns a pair whose first and second components are updates to $h$ and $ht$, respectively, that express the result of automatic maintenance of all rule sets  in heap $h$ and heap type map $ht$.  For each set of rules that needs to be maintained, it calls $\infupd$ to compute an update expressing the result of inference for that rule set, and uses function $\pi_1$, which select the first component of a tuple, to extract that update from the tuple returned by $\infupd$.  It combines the resulting updates using union, since the well-formedness restrictions on programs ensure that these updates have disjoint domains.
$\maintain$ uses recursion to repeatedly evaluate all rule sets until a fixed-point is reached\noticlp{, as discussed in Section \ref{sec-formal}}.  



\mypar{Notes}
The transition rules enforce the invariant that each non-local base predicate is either uninitialized or its value is a set.  Inference treats uninitialized variables used as base predicates as empty sets.  This is consistent with the semantics of Datalog and Prolog, which treats predicates for which no information has been supplied as false for all arguments.  
This principle is realized implicitly in the set comprehensions defining ${\it facts}_B$ and ${\it facts}_L$ in $\infupd$: the resulting sets do not contain any facts for those base predicates.  This principle applies whenever an uninitialized field is encountered in the sequence of field dereferences used to read the value of a base predicate.
 
The transition rules include premises that check for run-time errors; in case of an error, the premise is false, and evaluation is stuck.  Examples of such errors include trying to select a component from a value that is not a tuple, invoke a non-existent method of an object, read the value of a non-existent (uninitialized) field of an object, assign a value to a derived predicate using an assignment statement, or assign a non-set value to a base predicate.  The transition rules check for this error at updates to fields of instances of all classes---not only classes that define rule sets---because base predicates may contain multiple field dereferences.

Transition rules for methods of pre-defined classes \kwtt{set} and \kwtt{sequence} are similar in style, so only one representative example is given, for {\tt \kw{set}.\kw{add}}.  Note that $\maintain$ needs to be called only in transition rules for methods of \kwtt{set} that update the content of the set. 

The transition rule for invoking a method in a user-defined class executes a copy of the method body $s$ that has been instantiated by substituting argument values for parameters.

The transition rule for an explicit call to \kwtt{infer} on a rule set $rs$ with class scope instantiates $rs$ using the target object $a$ for \kwtt{self} and values given by keyword arguments for local variables, calls $\infupd$ to evaluate the instantiated rule set, and calls $\maintain$ to determine the effects of automatic maintenance.  Note that 
$\theta$ is an update to the heap that updates the values of non-local derived predicates of $\rs$; 
$\result$ maps each derived predicate of $R$ to its value;
$\theta_{\it QNL}$ and $\theta_{\it QL}$ are updates to the heap that together update the values of $a_1.f_1,\ldots,a_n.f_n$ to contain the query results, with the former handling queries of non-local derived predicates, and the latter handling queries of local derived predicates; 
and $\thetaht$ is an update to the heap type map that updates the types of addresses containing sets created by this call to \kwtt{infer}.


\begin{figure}[tbp]
\setstretch{0.96}
\begin{displaymath}
\begin{array}{@{}l@{}}
\commentS{context rule for expressions}\\
\dfrac{h,ht \vdash e \ra e'}{\tuple{C[e], h, ht} \ra \tuple{C[e'], h, ht}}\\
\\
\commentS{context rule for statements}\\
\dfrac{\tuple{s, h, ht} \ra \tuple{s', h', ht'}}{\tuple{C[s], h, ht} \ra \tuple{C[s'], h', ht'}}\\
\\
\commentS{field assignment}\\
\tuple{a{\tt .}f \;{\tt :=}\; v, h, ht}\\
{} \ra \tuple{\kwtt{skip}, h'\sqcup\theta, ht\sqcup\thetaht}\\
\sci{} \IF \legalassign(ht,a,f) \land h' = h[a \mapsto h(a)[f \mapsto v]] \land \allBaseAreSets(h',ht) \\
\sci {} \land \tuple{\theta,\thetaht}=\maintain(h',ht)\\
\\
\commentS{object creation}\\
\tuple{a{\tt .}f \;{\tt :=}\; \kwtt{new}~c, h, ht}\\
{}\ra \tuple{\kwtt{skip}, h'\sqcup\theta, ht'\sqcup\thetaht}\\
\sci \IF a'\not\in\dom(ht) \land a' \in \Address \land \legalassign(ht,a,f)\\
\sci {} \land ht' = ht\,[a' \mapsto c] \land h' =  h[a \mapsto h(a)[f \mapsto a'], a' \mapsto \new(c)] \land \allBaseAreSets(h',ht')\\
\sci {} \land \tuple{\theta,\thetaht}=\maintain(h',ht')
\\
\\
\commentS{sequential composition}\\
\tuple{{\tt \kw{skip};}\, s, h, ht} \ra
\tuple{s, h, ht}\\
\\
\commentS{conditional statement}\\
\tuple{\kwtt{if}~{\tt \kw{True}:}~s_1~{\tt \kw{else}:}~s_2, h, ht} \ra
\tuple{s_1, h, ht}\\
\\
\tuple{\kwtt{if}~{\tt \kw{False}:}~s_1~{\tt \kw{else}:}~s_2, h, ht} \ra
\tuple{s_2, h, ht}\\
\\
\commentS{for loop}\\
\tuple{\kwtt{for}~x~\kwtt{in}~a{\tt :}~s, h, ht}\\
{} \ra
\tuple{\kwtt{for}~x~\kwtt{inTuple}~{\tt (}v_1,\ldots,v_n{\tt ):}~s, h, ht}\\
\sci \IF (ht(a)=\kwtt{sequence} \land h(a)=\seq{v_1,\ldots,v_n})\\
\sci {} \lor
    (ht(a)=\kwtt{set} \land \seq{v_1,\ldots,v_n}~\mbox{is a linearization of}~h(a))\\
\\
\tuple{\kwtt{for}~x~\kwtt{inTuple}~{\tt (}v_1,\ldots,v_n{\tt ):}~s, h, ht}\\
{}\ra \tuple{s[x\mapsto v_1]; \kwtt{for}~x~\kwtt{inTuple}~{\tt (}v_2,\ldots,v_n{\tt ):}~s, h, ht}\\
\\
\tuple{\kwtt{for}~x~\kwtt{inTuple}~{\tt ():}~s, h, ht} \ra
\tuple{\kwtt{skip}, h, ht}\\
\\
\commentS{while loop}\\
\tuple{\kwtt{while}~e{\tt :}~s, h, ht} \\
{} \ra
\tuple{\kwtt{if}~e{\tt :}~{\tt (}s{\tt ;}~\kwtt{while}~e{\tt :}~s{\tt )}~{\tt \kw{else}:}~\kwtt{skip}, h, ht}
\end{array}
\end{displaymath}
  \caption{Transition relation for statements, Part 1.}
  \label{fig:transition-one}
\end{figure}

\begin{figure}[tbp]
\begin{displaymath}
\begin{array}{@{}l@{}}
\commentS{invoke method in pre-defined class (example)}\\
\tuple{a{\tt .add(}v_1{\tt )}, h, ht}\ra
\tuple{\kwtt{skip}, h'\sqcup\theta, ht\sqcup\thetaht}\\
\sci \IF ht(a)=\kwtt{set} \land \nexists \agv.f\in\derivedG:\; a=h(\agv)(f)\\
\sci {} \land \nexists a'\in\dom(ht), \kwtt{self}.f \in \nlderived(ht(a')):\; a = h(a')(f)\\
\sci {} \land h' = h[a \mapsto h(a)\union\set{v_1}] \land \tuple{\theta,\thetaht}=\maintain(h',ht)\\
\\
\commentS{invoke method in user-defined class}\\
\tuple{a{\tt .}m(v_1,\ldots,v_n), h, ht}\\
{} \ra
\tuple{s[\kwtt{self}\mapsto a, x_1\mapsto v_1, \ldots, x_n\mapsto v_n], h, ht}\\
\sci\IF \methodDef(ht(a), m, \kwtt{def}~m{\tt (}x_1,\ldots,x_n{\tt )}~s)\\
\\
\commentS{invoke \kwtt{infer} on a rule set defined in class scope}\\
\ltuple a_1.f_1,\ldots,a_n.f_n := a{\tt .}\kwtt{infer(}q_1,\ldots,q_n, x_1=v_1, \ldots, x_k=v_k,\kwtt{rules}{\tt =}rs{\tt )}, h, ht\rtuple\\
\sci {} \ra \tuple{\kwtt{skip}, h'\sqcup\theta', ht\sqcup\thetaht\sqcup\thetaht'}\\
\sci\IF rs \in \rulesets(ht(a))\\
\sci 
\begin{array}{@{}l@{}}
{} \land \forall i \in \set{1..n}: \legalassign(ht,a_i,f_i)\\
{} \land (\forall i \in \set{1..k}: v_i\in\dom(ht) \land ht(v_i)=\kwtt{set})\\
{} \land \args = \set{x_i \mapsto v_i \;|\; i \in \set{1..k}}\\
{} \land \tuple{\theta,\result} = \infupd(h, ht, \rs, a, \args)\\
{} \land \theta_{\it QNL} = \UNION_{i \in \set{1..n} \mbox{ s.t. $q_i$ is a non-local predicate } \kwtt{self}.f} \set{a_i \mapsto  h(a_i)[f_i \mapsto (h\sqcup\theta)(a)(f)]}\\
{} \land \theta_{\it QL} = \UNION_{i \in \set{1..n} \mbox{ s.t. $q_i$ is a local predicate}} 
\{\begin{array}[t]{@{}l@{}}
  a_i \mapsto  h(a_i)[f_i \mapsto \newaddr(\rs,q_i,h)],\\
 \newaddr(\rs,q_i,h) \mapsto \result(q_i)\}
 \end{array}\\
{} \land \thetaht = 
\begin{array}[t]{@{}l@{}}
  \set{a \mapsto \kwtt{set} \;|\; a\in\dom(\theta)  \land \theta(a) \subseteq \Val}\\
  {} \union
  \set{a \mapsto \kwtt{set} \;|\; a\in\dom(\theta_{\it QL})  \land \theta_{\it QL}(a) \subseteq \Val}
\end{array}\\
{} \land h' = h\sqcup\theta\sqcup\theta_{\it QNL}\sqcup\theta_{\it QL}\sqcup\theta_U\\
{} \land \allBaseAreSets(h', ht\sqcup\thetaht)\\
{} \land \tuple{\theta',\thetaht'}=\maintain(h',ht\sqcup\thetaht)
\end{array}
\end{array}
\end{displaymath}
  \caption{Transition relation for statements, Part 2.}
  \label{fig:transition-two}
\end{figure}

\mypar{Executions}
An execution is a sequence of transitions $\sigma_0 \ra \sigma_1 \ra
\sigma_2 \ra \cdots$ such that $\sigma_0$ is the initial state of the program, given by $\sigma_0 = \tuple{s_0, \set{\agv \mapsto \emptyfn}, \set{\agv \mapsto \classgv}}$, where $s_0$ is the top-level statement that appears in the program after the rule set definitions and class definitions, and $\agv$ is the address of the object introduced by the transformation that eliminates global variables (see Section \ref{sec:translation}).

Execution of a program may eventually (1) terminate (i.e., the statement in the first component of the state becomes \kwtt{skip}, meaning that there is nothing left to do), (2) get stuck (i.e., the statement is not \kwtt{skip}, and the process has no enabled transitions, meaning an error in the program), or (3) run forever (due to an infinite loop or infinite recursion).  



}{}
\oops{

\input{experiment}
\input{supplementary.bbl}
}

\end{document}